\documentstyle{mn}
\input psfig.tex
\newcommand{\ltsima} {$\; \buildrel < \over \sim \;$}
\newcommand{\gtsima} {$\; \buildrel > \over \sim \;$}
\newcommand{\lta} {\lower.5ex\hbox{\ltsima}}
\newcommand{\gta} {\lower.5ex\hbox{\gtsima}}

\def\refitem{\par\parskip 0pt\noindent\hangindent 20pt}

\title[Unification of blazars]
{A theoretical unifying scheme for gamma-ray bright blazars}

\author[G. Ghisellini, A. Celotti, G. Fossati, L. Maraschi, A. Comastri]
{G. Ghisellini$^1$ \thanks{E-mail: {\tt gabriele@merate.mi.astro.it}},
A. Celotti$^{2,3}$ G. Fossati$^2$, L. Maraschi$^1$ and A. Comastri$^4$\\
$^1$ Osservatorio Astronomico di Brera, via Bianchi 46, I-22055 Merate, Italy\\
$^2$ S.I.S.S.A., via Beirut 2-4, 34014 Trieste, Italy \\
$^3$ Institute of Astronomy, Madingley Road, Cambridge CB3 0HA \\
$^4$ Osservatorio Astronomico di Bologna, via Zamboni 33, I-40126, Bologna, Italy\\}

\date{Received ***; in original form ***}

\begin{document}
\maketitle

\begin{abstract} 
The phenomenology of $\gamma$--ray bright blazars can be accounted 
for by a sequence in the
source power and intensity of the diffuse radiation field surrounding
the relativistic jet.  Correspondingly, the equilibrium particle
distribution peaks at different energies.  This leads to a trend in
the observed properties: an increase of the observed power corresponds
to: 1) a decrease in the frequencies of the synchrotron and inverse
Compton peaks; 2) an increase in the ratio of the powers of the high
and low energy spectral components.  Objects along this sequence would
be observationally classified respectively as high frequency BL Lac
objects, low frequency BL Lac objects, highly polarized quasars and
lowly polarized quasars.
 
The proposed scheme is based on the correlations among the physical
parameters derived in the present paper by applying to 51
$\gamma$--ray loud blazars two of the most accepted scenarios for the
broad band emission of blazars, namely the synchrotron self--Compton
and external Compton models, and explains the observational trends
presented by Fossati et al. (1998) in a companion paper, dealing with 
the spectral energy distributions of all blazars.
This gives us confidence that our scheme applies to all blazars as a class.
\end{abstract}


\section{Introduction}
Among Active Galactic Nuclei (AGN) blazars represent the most extreme
and powerful sources.  The fundamental property characterizing blazars
is their beamed continuum, due to plasma moving relativistically along
the line of sight.

This scenario seems to apply to objects with somewhat different
observational properties leading to different
classifications/definitions.  Objects with significant emission line
equivalent widths are usually found as flat spectrum radio quasars
(FSRQ).  Objects without emission lines (EW $<$ 5 \AA) are classified
as BL Lac objects.  Different flavors of BL Lac objects have been
found in radio and X-ray surveys.  These also correspond to
differences in the overall spectral energy distributions (SED)
(see. e.g. Padovani \& Giommi 1995), which have been interpreted
either as due to orientation (Ghisellini \& Maraschi 1989, Urry \&
Padovani 1995), or as intrinsic (Padovani \& Giommi 1995).
Nevertheless, while different sub--classes have different average
properties, the actual distinction among them is certainly fuzzy and
so far several sources have shown intermediate behavior.  In fact
arguments for a substantial `continuity' in the continuum spectral
properties leading to adopt the blazar denomination as including both,
BL Lacs as well as FSRQs, have been recently re--proposed by Maraschi
et al. (1995), Sambruna et al. (1996) and Fossati et al. (1997).

The recent discoveries of about $\sim$60 blazars emitting in the
$\gamma$--ray band, by EGRET on board the Compton Gamma--Ray
Observatory (CGRO) (Fichtel et al. 1994; von Montigny et al. 1995;
Thompson et al. 1995; Mattox et al. 1997) and of a few BL Lac objects
by WHIPPLE and HEGRA (Weekes et al. 1996, Petry et al. 1996), have
revealed that the bulk of their radiative output is emitted in the
$\gamma$--ray range, thus allowing us to discuss for the first time
the characteristics of blazars knowing their total emission output and
their entire SED. At the same time these observations have raised
again the question as to whether and how the various subclasses differ
in their $\gamma$--ray properties.

For a deeper understanding of the fundamental mechanisms at work in
these sources it is crucial to address the questions: within the
blazar phenomenon which is the physical origin of the difference among
BL Lacs and even more broadly between BL Lacs and FSRQ?  Is it
possible to identify continuity among them, with a limited number of
physical properties determining the observational characteristics of
all blazars?


We address these issues from two sides: a purely observational
approach (Fossati et al. 1998) based on complete sub--samples of
blazars and (here) a more theoretical approach based on modeling
individually the SEDs of all the $\gamma$--ray sources with sufficient
available data to constrain their physical parameters.  This allows us
to derive trends between the physical quantities underlying the
correlations between the observed ones.

Several models, still in competition, have been proposed to explain
the $\gamma$--ray emission and the overall SED of blazars.  Mannheim
(1993) proposed that shock--accelerated electrons and protons give
origin to two different populations of particles (electrons and
electron--positron pairs), responsible of the entire SED through
synchrotron emission.  In an alternative widely adopted scenario, a
single population of electrons is supposed to radiate from the far IR
(or even radio) to the UV--soft X--rays by the synchrotron mechanism,
and at higher frequencies by the inverse Compton process.  In general
the observed SEDs require curved spectra steepening at higher
frequencies, for both the synchrotron and inverse Compton components.
In the $\nu F_\nu$ representation of the SEDs each component shows
then a peak that in the following will be referred to respectively as
the synchrotron and inverse Compton peak.  Specific models differ in
the adopted geometry (one--zone homogeneous models or inhomogeneous
jet models), and in the nature of the target photons which are
up-scattered in energy by the inverse Compton process.  The target
photons could be either synchrotron photons (Maraschi, Ghisellini \&
Celotti 1992; Bloom \& Marscher 1993; Ghisellini \& Maraschi 1994), or
be produced in the accretion disk (Dermer \& Schlickeiser 1993), or in
the broad line region (BLR).  The BLR itself can be either illuminated
by the disk (Sikora, Begelman \& Rees 1994; Blandford 1993; Blandford
\& Levinson 1995), or self--illuminated by the jet (Ghisellini \&
Madau 1996).  Finally, target photons could be produced by a dusty
torus surrounding the blazar nucleus (Wagner et al. 1995).  All these
different scenarios have been tested on specific sources, but often
more than one model can reproduce the same data with similar accuracy
(see von Montigny et al. 1997 for 3C 273; Ghisellini, Maraschi \&
Dondi 1996 for 3C 279; Comastri et al. 1997 for 0836+710).

Here we examine only two of the leading pictures, namely the
synchrotron self--Compton (SSC) and the `external Compton' (EC) model,
in which the main contribution to the target photons is produced
outside the $\gamma$--ray emitting region, even if some contribution
from the SSC component is always present.
Therefore, in the following, with the term ``EC" we mean 
a model in which both the SSC and the EC contributions to the
high energy spectrum are considered, while external photons are
completely neglected in the SSC model.
The SSC and the EC models are applied to all sources with sufficient 
available data to constrain the models themselves.

Through a search in the literature the multiwavelength overall spectra
of 51 $\gamma$--loud blazars have been assembled.  Even if the vast
majority of the data are not simultaneous and the sample is not
complete in any respect, they provide a useful template of the SED of
different classes of $\gamma$--loud blazars.  While the
non--simultaneity of the data (except for a few sources) precludes
from deriving strong conclusions about specific objects, their large
number allows us to study trends in the physical parameters of the
models and possible correlations among them and with the observed
spectral characteristics of different sub--classes of blazars.

In Section 2 the sample of sources is defined, while in Section 3 we
describe the two adopted models, the computing procedures and the
`fit' criteria.  The results are presented in Section 4 and discussed
in Section 5.

\section{The sample}

Two pieces of information have been considered essential for including
a source in the present sample:

\begin{enumerate}

\item either detection and estimate of the $\gamma$--ray spectral
slope in the EGRET band $or$ detection by the WHIPPLE observatory;

\item measured (or lower limit on) redshift.

\end{enumerate}
For all the sources satisfying the above criteria sufficient
information at lower frequencies could be found so that the location
of both the synchrotron and inverse Compton peaks and the luminosity
of each source could be estimated.  The resulting 51 sources are
listed in Table~1 in the Appendix, together with their redshift,
classification and the list of references relative to the data
plotted in Fig.~1a-f.  The sample includes 14 BL Lac objects and 37
quasars.  Among quasars, all core dominated radio sources with flat
radio spectra, 16 are HPQ (highly polarized: optical polarization $>$3
per cent), 16 are LPQ (lowly polarized), while for the remaining 5,
labeled NP, polarization measurements were not found.

BL Lacs can be divided in two sub--classes with different broad band
spectra according to their radio--X--ray spectral index $\alpha_{\rm RX}$ 
(Padovani \& Giommi 1995, Fossati et al. 1998).  
In fact there is a close correlation
between the value of $\alpha_{\rm RX}$ and the energy of the
synchrotron emission peak: for $\alpha_{\rm RX}>0.75$ this is in the
IR--optical (LBL: low frequency BL Lac), otherwise in the UV--soft
X--ray band (HBL: high frequency BL Lac).  According to this
definition, we have 10 LBL and 4 HBL.

\begin{figure*}
\psfig{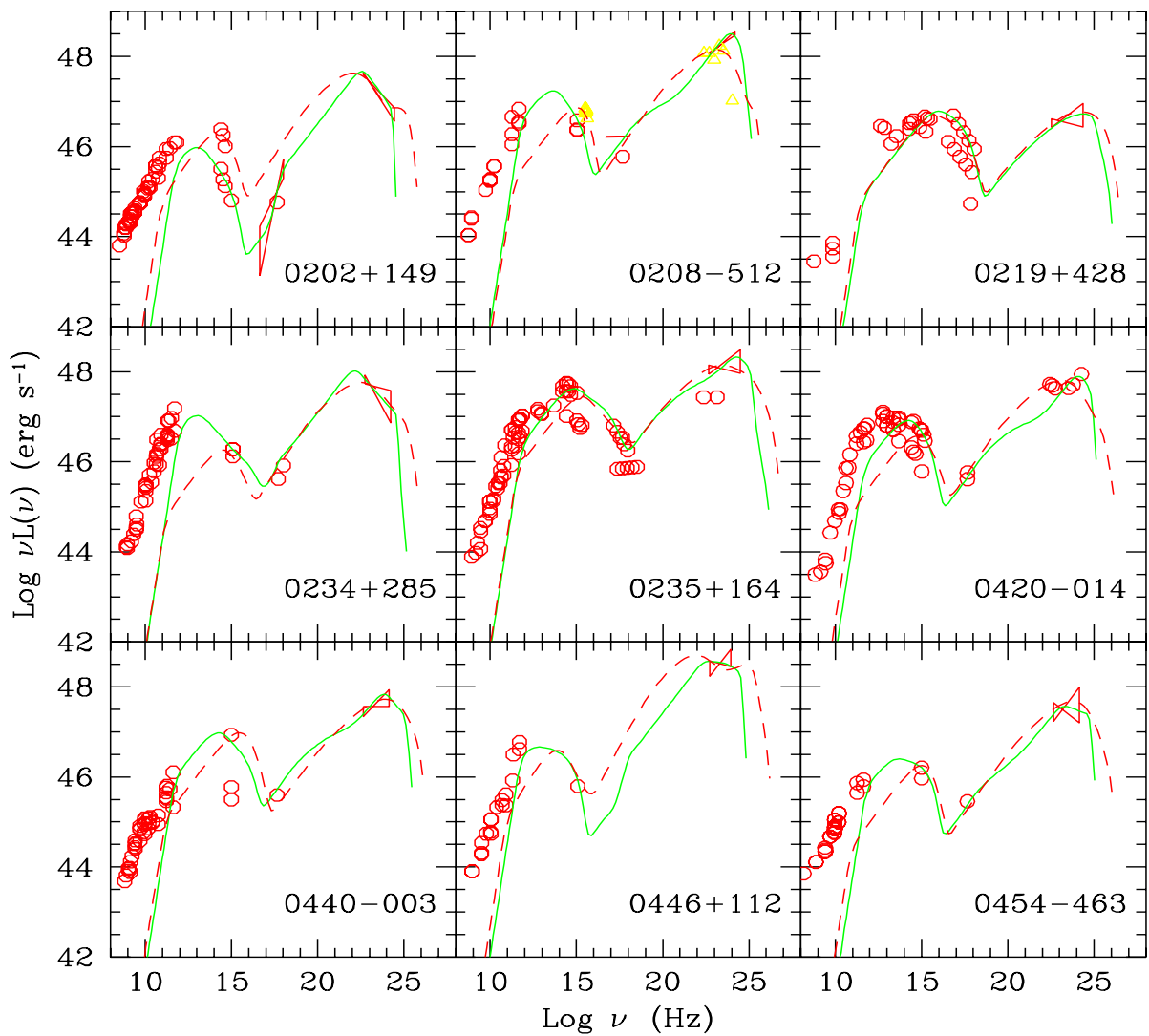}
\caption[h]{{a}. Spectral energy distributions (in $\nu L(\nu)$) of
the 51 $\gamma$--ray loud sources. The broad band spectra have been
assembled from data in the literature (the complete list of references
is given in Table~1). In parenthesis the rescaling factors used for
graphical purpose are indicated.  SED from the SSC and EC models are
superposed to the data, as dashed and solid line, respectively. The
model parameters are reported in Table~2, in the Appendix.  For
0716+714 a redshift $z=0.3$ has been adopted.}
\end{figure*}
\setcounter{figure}{0}
\begin{figure*}
\psfig{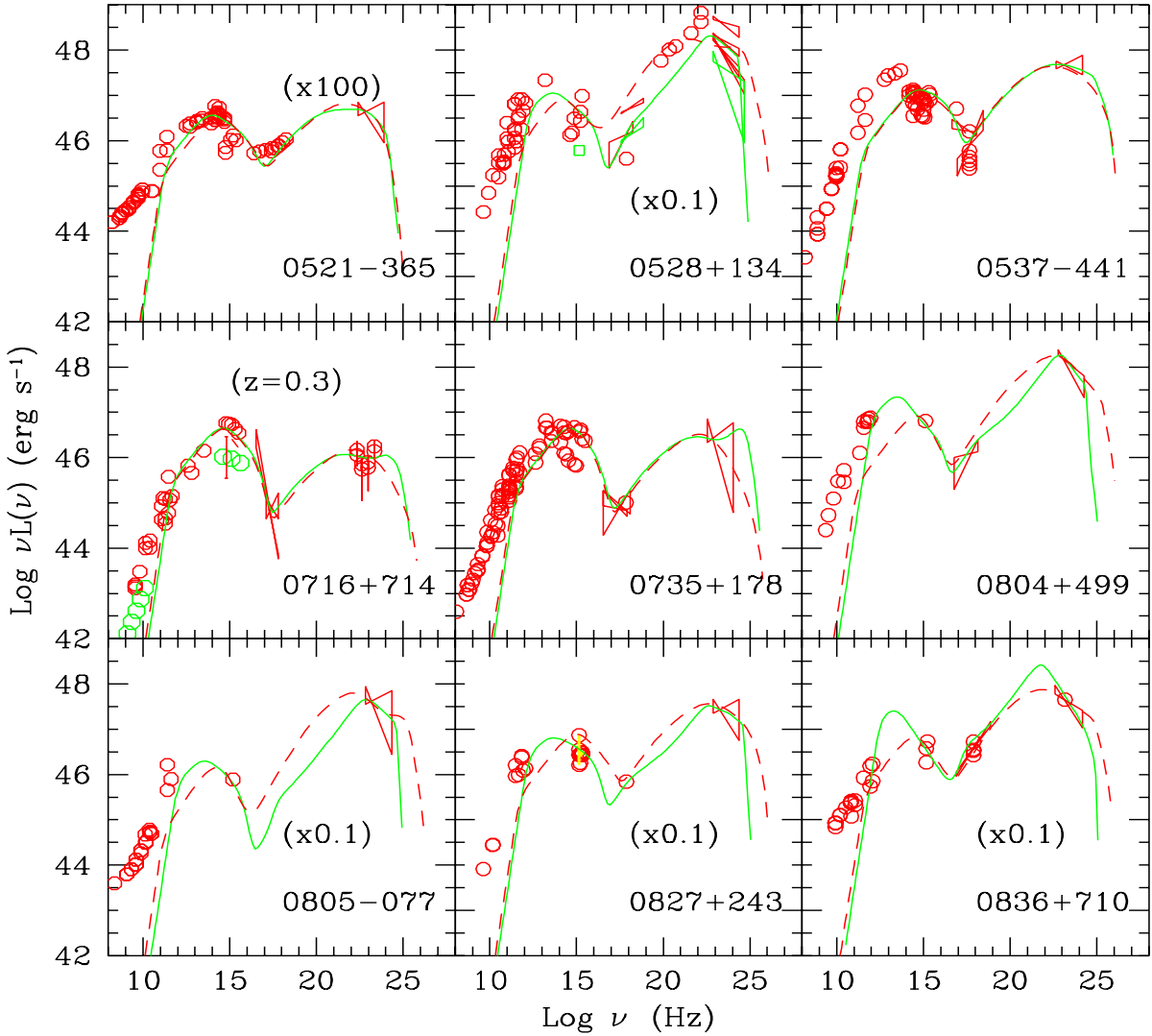}
\caption[h]{{b} Same as Fig. 1a}
\end{figure*}
\setcounter{figure}{0}
\begin{figure*}
\psfig{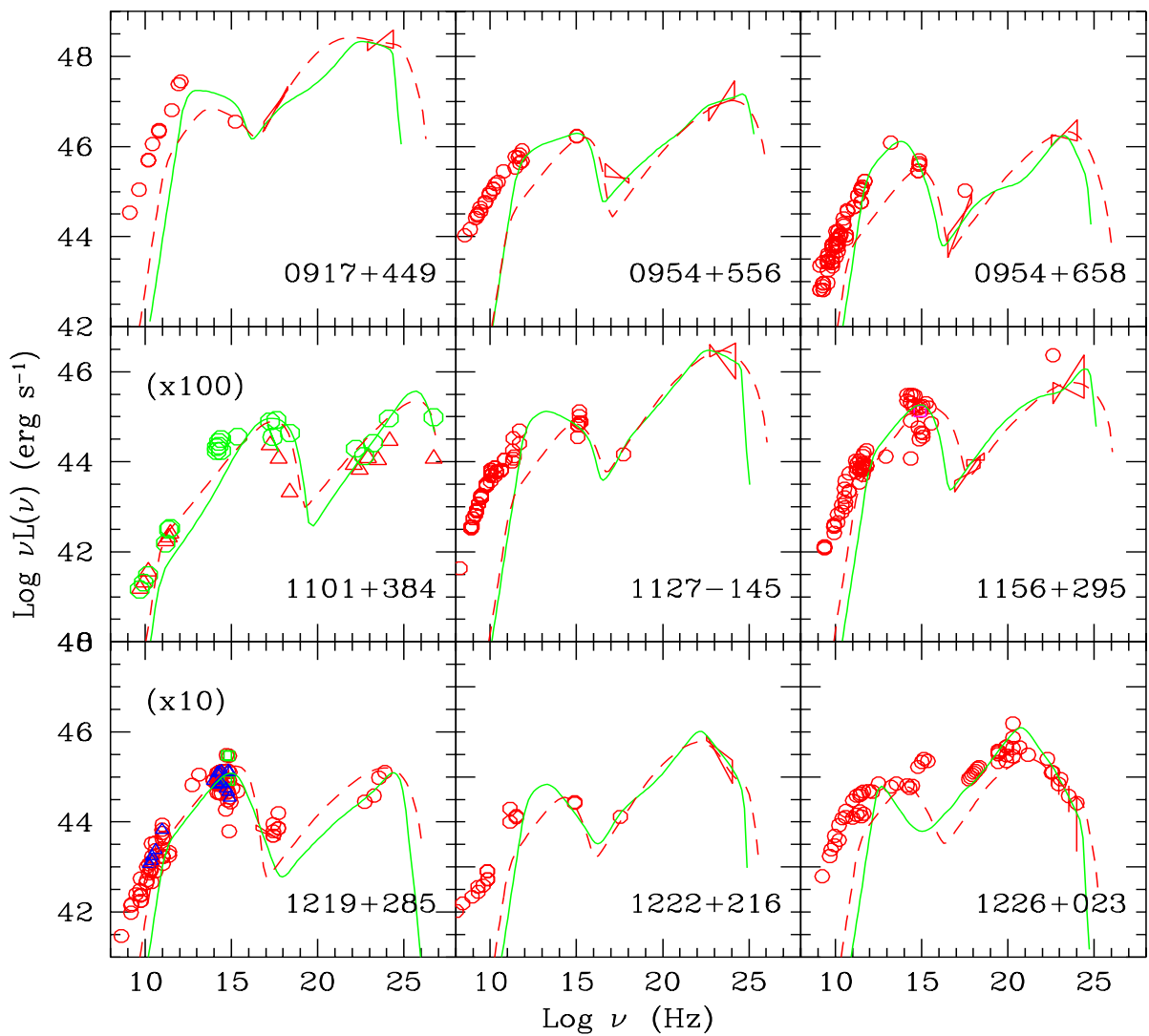}
\caption[h]{{c} Same as Fig. 1a}
\end{figure*}
\setcounter{figure}{0}
\begin{figure*}
\psfig{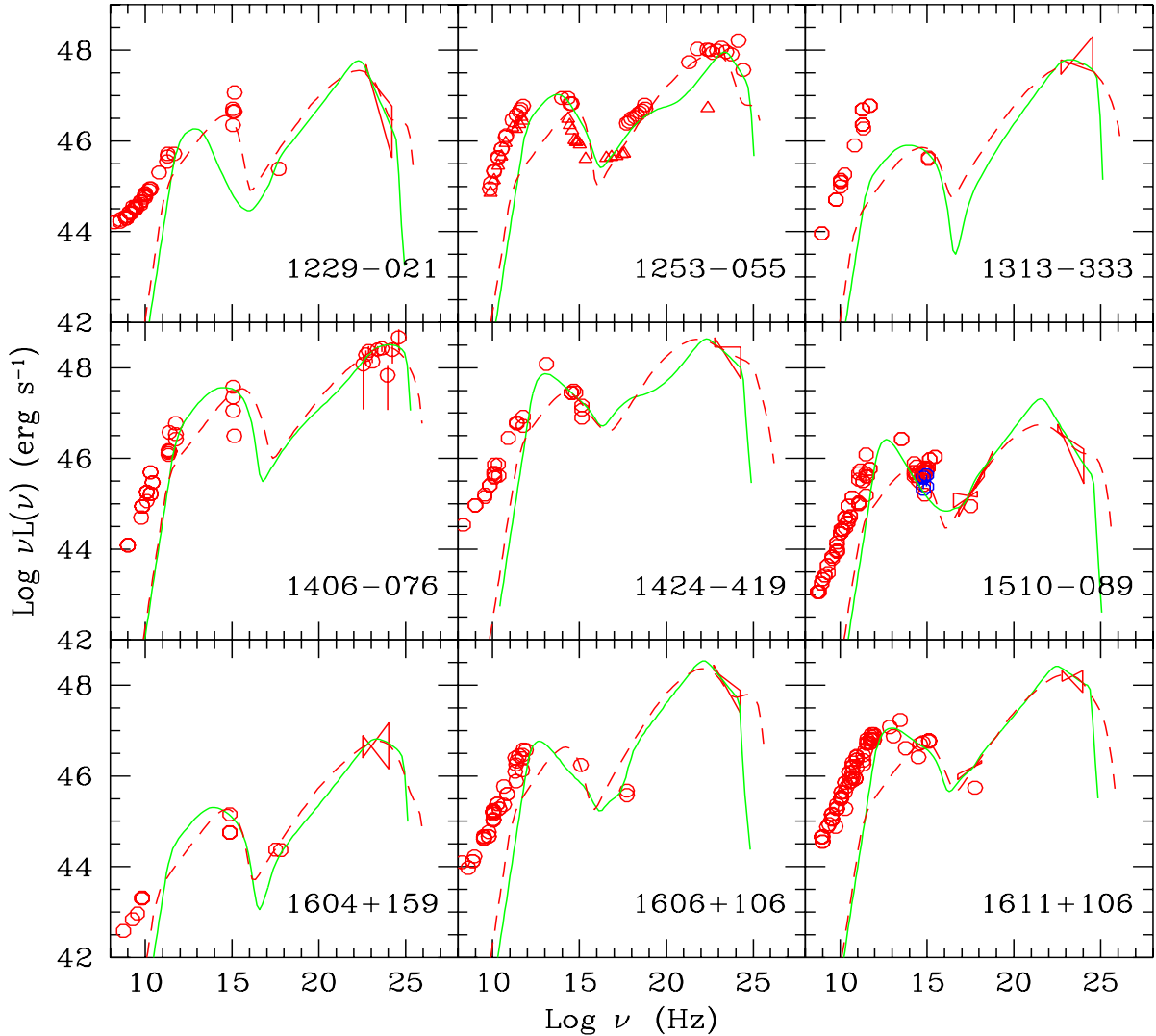}
\caption[h]{{d} Same as Fig. 1a}
\end{figure*}
\setcounter{figure}{0}
\begin{figure*}
\psfig{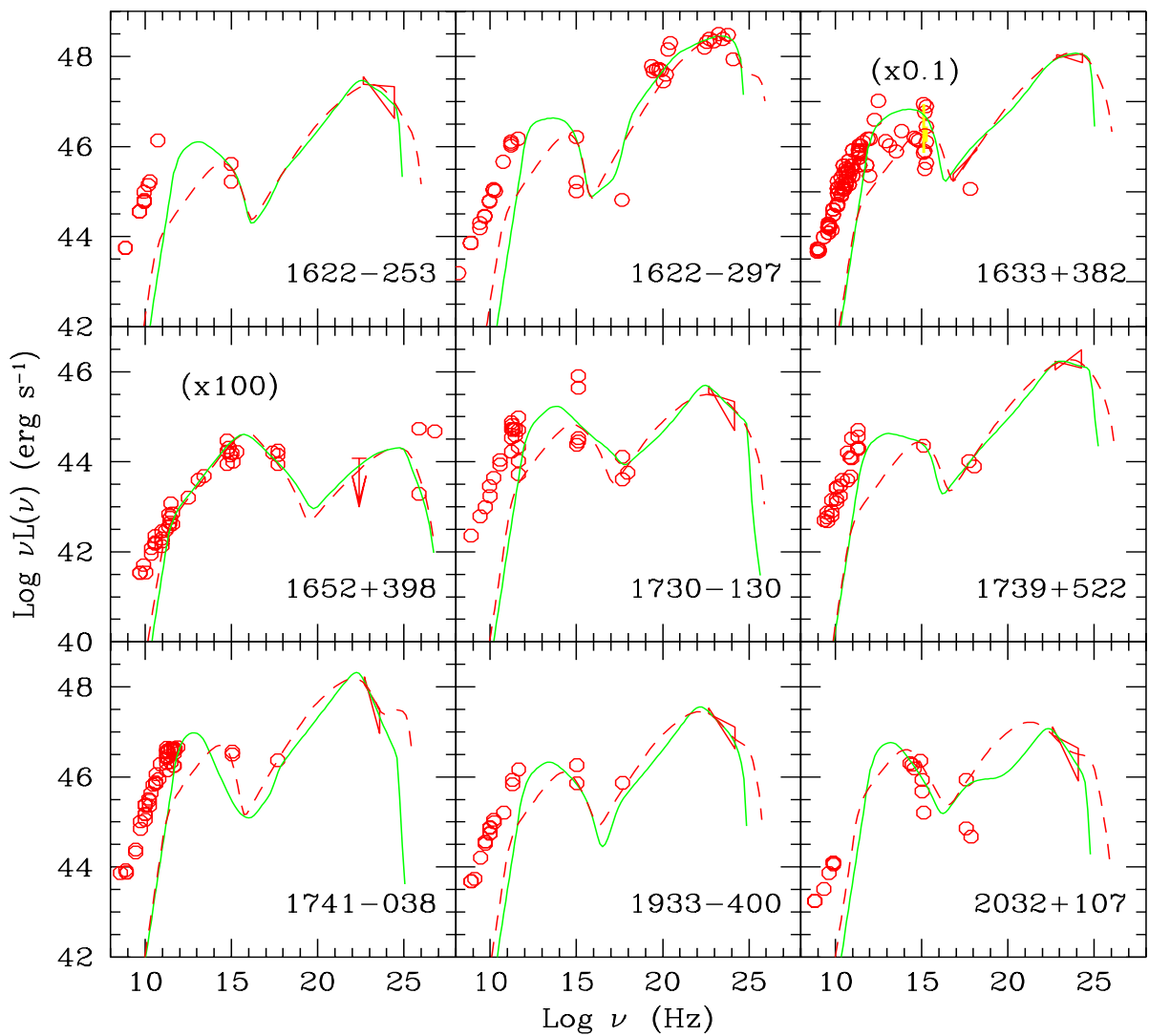}
\caption[h]{{e} Same as Fig. 1a}
\end{figure*}
\setcounter{figure}{0}
\begin{figure*}
\psfig{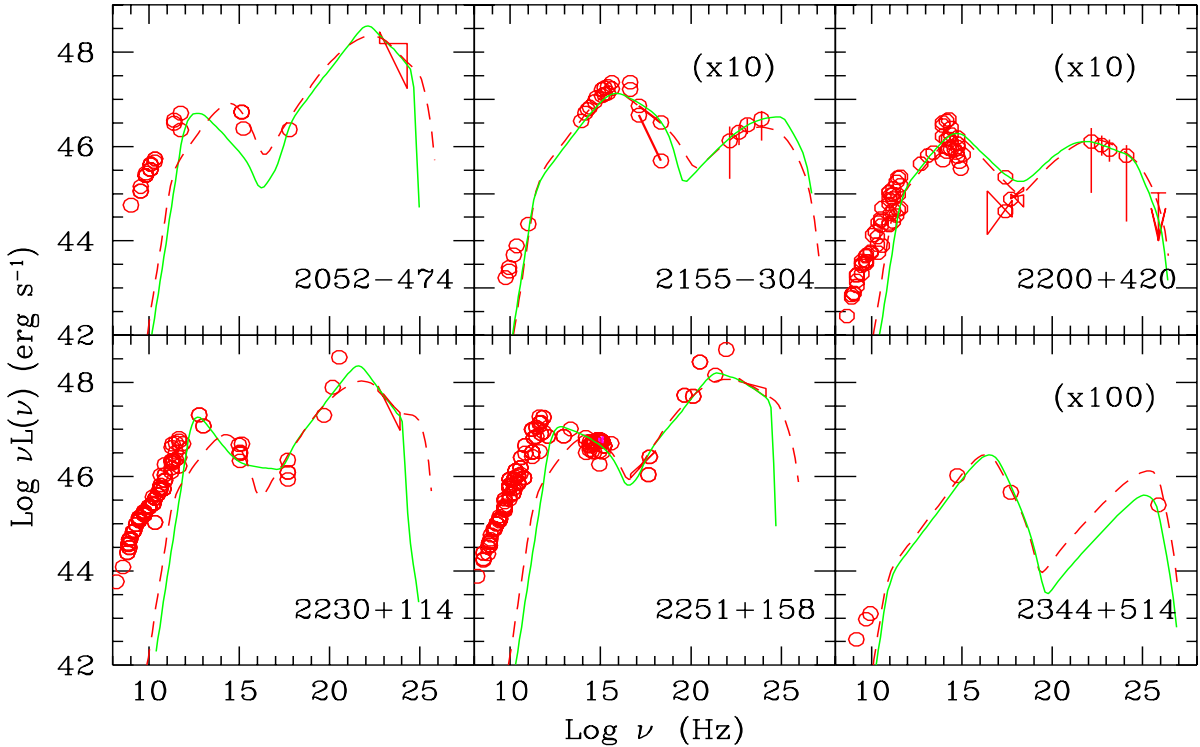}
\vskip -5 truecm
\caption[h]{{f} Same as Fig. 1a}
\end{figure*}

Note that in our list there are two sources detected by WHIPPLE but
not by EGRET, i.e. the two HBL objects Mkn 501 (1652+398) and 1ES
2344+514.

In Fig.~1a-f, the overall SEDs of all the blazars  listed in Table~1
are plotted.  Frequencies and luminosities are in the rest frame of
the source and are calculated assuming cosmological parameters
$H_0=50$ km s$^{-1}$ Mpc$^{-1}$ and $q_0=0.5$.  Fluxes have been
dereddened using the absorption values reported in the NED database.

\section{The models}
\subsection{General assumptions}

The emitting region is assumed to be a sphere (blob) of 
constant radius $R$,
with a homogeneous and tangled magnetic field $B$.  Throughout the
source relativistic electrons are continuously injected at a rate
$Q(\gamma)$ [cm$^{-3}$ s$^{-1}$], corresponding to a luminosity
$L_{\rm inj}$ and a compactness $\ell_{\rm inj} \equiv L_{\rm
inj}\sigma_{\rm T}/(R\, m_{\rm e} c^3)$, where $\sigma_{\rm T}$ is the
Thomson scattering cross section.  This power is assumed to be
entirely converted into radiation.  The injected particles are
distributed in energy as a power--law of slope $s$ [$Q(\gamma)= 
Q_0 \gamma^{-s}$], between $\gamma_{\rm min}$ and $\gamma_{\rm max}$.

The blob moves with a bulk velocity $\beta c$, corresponding to a
Lorentz factor $\Gamma$, at an angle $\theta$ with respect to the line
of sight.  The Lorentz transformation of the specific intensity is
thus given by $I(\nu)=\delta^3 I^\prime(\nu/\delta)$, where
$\delta=[\Gamma(1-\beta\cos\theta)]^{-1}$ is the Doppler factor. For
simplicity (see below), we always assume $\theta\sim 1/\Gamma$,
resulting in $\delta\sim \Gamma$.  In the following of this section,
unless otherwise specified, all quantities are measured in the blob
comoving frame.

We consider a stationary situation, that is we determine the particle
equilibrium distribution and the spectrum of the emitted radiation
self--consistently, assuming that the timescale over which the
particles reach equilibrium is shorter than that over which the
injection mechanism changes. We neglect particle escape and adiabatic
expansion.

\subsection{The particle distribution}

The equilibrium particle distribution $N(\gamma)$ (cm$^{-3}$) is
determined by the stationary solution of the continuity equation
$$
N(\gamma) \, =\, { \int_\gamma^{\gamma_{\rm max}} [Q(\gamma)+P(\gamma)]
d\gamma \over \dot\gamma}
\eqno(1)
$$
where $\dot\gamma$ is the cooling term and $P(\gamma)$ is the rate of
electron--positron pair production.  The only important mechanism for
pair production is photon--photon collisions, the rate of which is
calculated according to the prescriptions given in e.g.  Ghisellini
(1989).

$\dot\gamma$ takes into account the following cooling mechanisms:

\begin{enumerate}

\item synchrotron emission: $m_{\rm e} c^2\dot\gamma_{\rm
s}=(4/3)\sigma_{\rm T} c \gamma^2 U_{\rm B}$, where $U_{\rm B} = 
B^2/(8\pi)$ is the magnetic energy density;

\item inverse Compton emission: $m_{\rm e} c^2\dot \gamma_{\rm C}=
(4/3)\sigma_{\rm T} c \gamma^2 U_{\rm r}$, where $U_{\rm r}$ is the
radiation energy density.  Since the radiation spectrum extends to
high energies, the scattering process has to be calculated by means of
the Klein--Nishina cross section.  For simplicity, we approximate it
with a step function equal to the Thomson cross section for
frequencies $x\equiv h\nu/(m_{\rm e}c^2) \le (3/4)/\gamma$, and zero
otherwise.  This implies that the radiation energy density effectively
involved in the inverse Compton cooling depends on the electron energy
$$
U_{\rm r}(\gamma) \, = \, m_{\rm e}c^2\int_0^{3/(4\gamma)}U(x)dx
\eqno(2)
$$

The continuity equation is solved numerically, with an iterative
approach, as described in Ghisellini (1989).  The numerical treatment
is necessary because of the high non--linearity of the processes
involved: $N(\gamma$) depends on the radiation spectrum (because of
the inverse Compton cooling term and the pair production rate), which
in turn is determined by $N(\gamma)$.

When the Klein-Nishina and pair production effects can be neglected,
the solution of equation 1 is trivial: 1) for injection indices $s>2$,
we have a broken power law: $N(\gamma)\propto \gamma^{-2}$ up to
$\gamma_{\rm min}$ and $N(\gamma)\propto \gamma^{-s-1}$ above.  In
this case $\gamma_{\rm min}$ can be identified with the crucial
parameter $\gamma_{\rm peak}$, i.e. the Lorentz factor of the
electrons emitting at the peaks of the synchrotron and inverse Compton
components; 2) if $1<s<2$ we have the same solutions for $N(\gamma)$,
but in this case $\gamma_{\rm min}<\gamma_{\rm peak}<\gamma_{\rm
max}$, since the spectral index of the radiation emitted by particles
above $\gamma_{\rm min}$ is flatter than unity; 3) if $s<1$ the lower
limit of the integral in equation 1 becomes unimportant, yielding
$N(\gamma)\propto \gamma^{-2}$ in the entire energy range, except for
$\gamma$ close to $\gamma_{\rm max}$.

Note that the assumption of constant radius and no escape
tends to overestimate the particle distribution at the lowest 
energies,
where these effects are potentially more important than radiative
cooling (if the overall compactness is much less than unity).
This has no effect on the synchrotron spectrum, which is
self--absorbed at low frequencies, and has no effect on the
observable SSC spectrum, mainly made by high energy electrons.
In the case of the EC model, instead, the X--ray spectrum
is made by the sum of the EC and SSC components, and therefore
the X-ray flux and spectrum can depend on the details of the low
energy particle distribution if the EC component dominates.
Then in these cases the calculated X--ray spectrum 
could be flatter than what derived here.

\end{enumerate}

\subsection{Target photons}

$U_{\rm r}(\gamma)$ includes the contribution from the radiation both
produced internally (by synchrotron and self--Compton emission) and
externally to the blob.

The latter is assumed to be distributed as a (diluted) blackbody,
peaking at a frequency $x_{\rm ext} \equiv h\nu_{\rm ext}/(m_{\rm e}c^2)$ 
between $5\times 10^{-5}$ and $2\times 10^{-4}$ (in the rest
frame of the blob). The exact value depends on the radiation mechanism
responsible for the external field and the bulk Lorentz factor of the
blob. The assumption of a blackbody spectral distribution is merely
for ease of calculation. For instance, in the case of external
radiation dominated by the broad emission line photons, an observer in
the comoving frame of the blob would see a complex spectrum, not
isotropic (blueshifted in the forward direction and redshifted in the
opposite one): even a single, monochromatic line would be transformed
into a peaked, but extended, spectrum.

Consequently, a peaked distribution can approximate the case of
externally produced photons distributed in lines, independently of the
origin of the photoionizing continuum: we can treat the cases of
disk--illuminated as well as jet--illuminated BLR.  On the other hand,
this assumption can mimic the effect of an external scattering medium
only if the illuminating continuum is narrowly distributed in
frequency (e.g. radiation produced by an accretion disk), but it is
not satisfactory for a scattering medium illuminated by the jet (which
produces a more extended spectrum).

For a direct comparison with the value of the compactness in injected
electrons, $\ell_{\rm inj}$, we assume that also the external
radiation can be characterized by an `effective compactness'
$\ell_{\rm ext}$, defined as
$$
\ell_{\rm ext}\, =\, 
{\sigma_{\rm T} R U_{\rm ext} \over m_e c^2}
\eqno(3)
$$
where $U_{\rm ext}$ is the radiation energy density (of the external
radiation) as seen in the comoving frame, and is therefore amplified
by a factor $\Gamma^2$ with respect to the same quantity measured in
the frame of the observer.

As already mentioned, this external field is not isotropic in the
comoving frame (see Dermer 1995).  However both for simplicity and
because of the uncertainty in the origin and therefore in the angular
distribution of the external radiation, we assume an isotropic pattern
for $\ell_{\rm ext}$ (in the comoving frame).  With this approximation
also the inverse Compton radiation is isotropically distributed in
this frame, and subject to the same Lorentz transformation as the
synchrotron and self--Compton emission.

The uncertainty related to the latter assumption can be estimated 
comparing the two extreme cases of the Compton flux emitted 
assuming (in the comoving frame) an isotropic seed photon 
distribution and the case of soft photons distributed 
only along the jet axis.
Assume for simplicity that in both cases the seed photons
are monochromatic, at the frequency $\nu_0^\prime$.
The total power emitted by an electron of energy $\gamma  m_ec^2$ is  
$P=\sigma_T c 
U^\prime_{\rm rad}[\gamma^2\int (1-\beta\cos\phi)^2 d\Omega/(4\pi) -1]$
(see e.g. Rybicky \& Lightman 1979), where $\phi$ is the angle
between the incoming photon and the electron velocity vector and
$U^\prime_{\rm rad}$ is the energy density of the seed photons.
The viewing angle $\theta=1/\Gamma$ corresponds to the aberrated angle
$\theta^\prime=90^\circ$: at this angle, the power received in the
isotropic case is $P_{\rm iso} = (4/3)\sigma_T c U^\prime_{\rm rad}$,
while the power received in the monodirectional case ($\phi=90^\circ)$
is $P_{\rm mono} = \sigma_T c U^\prime_{\rm rad}(\gamma^2/2-1)$.
The ratio $P_{\rm mono}/P_{\rm iso}$
for large $\gamma$ is therefore equal to $3/8$.
The corresponding ratio between the scattered frequencies is
equal to 3/2.

\subsection{Observational constraints}

We require that the model parameters, besides giving a good
description of the broad band SED, also satisfy additional
constraints, regarding the observed variability timescales 
$t_{\rm var}$ and the amount of Doppler boosting.  In fact, as commonly
observed for the optical--UV and $\gamma$--ray emission of blazars,
the minimum variability timescale must be as short as a day, or a
fraction of a day.  This corresponds to demand
$$
R \lta c\ t_{\rm var} {\delta\over 1+z}
\eqno(4)
$$
with $t_{\rm var}\sim$1 day.  

Furthermore, the Doppler factor is constrained not to exceed a value
of 20--25, to be consistent with the observed superluminal speeds.

A third requirement, which is not imposed a priori but has to be
satisfied in all cases, concerns the amount of pair production.  As
discussed in Ghisellini \& Madau (1996), the $\gamma$--ray emitting
region $must$ be thin to the high energy radiation, otherwise it
inevitably leads to overproduction of X--rays. The line of the
argument is as follows: if a substantial fraction of the power emitted
in the $\gamma$--ray band gets absorbed in photon--photon collisions,
the pairs created (which are relativistic) radiate their energy in
other bands by synchrotron and inverse Compton emission. In particular
this results in a copious production of X--rays, with a luminosity of
the same order as that in $\gamma$--rays. Since the importance of the
pair production process is measured by the compactness, the
transparency requirement translates into an upper limit to the allowed
values of $\ell_{\rm inj}\lta 1$ (see Dondi \& Ghisellini 1995).

\begin{figure*}
\psfig{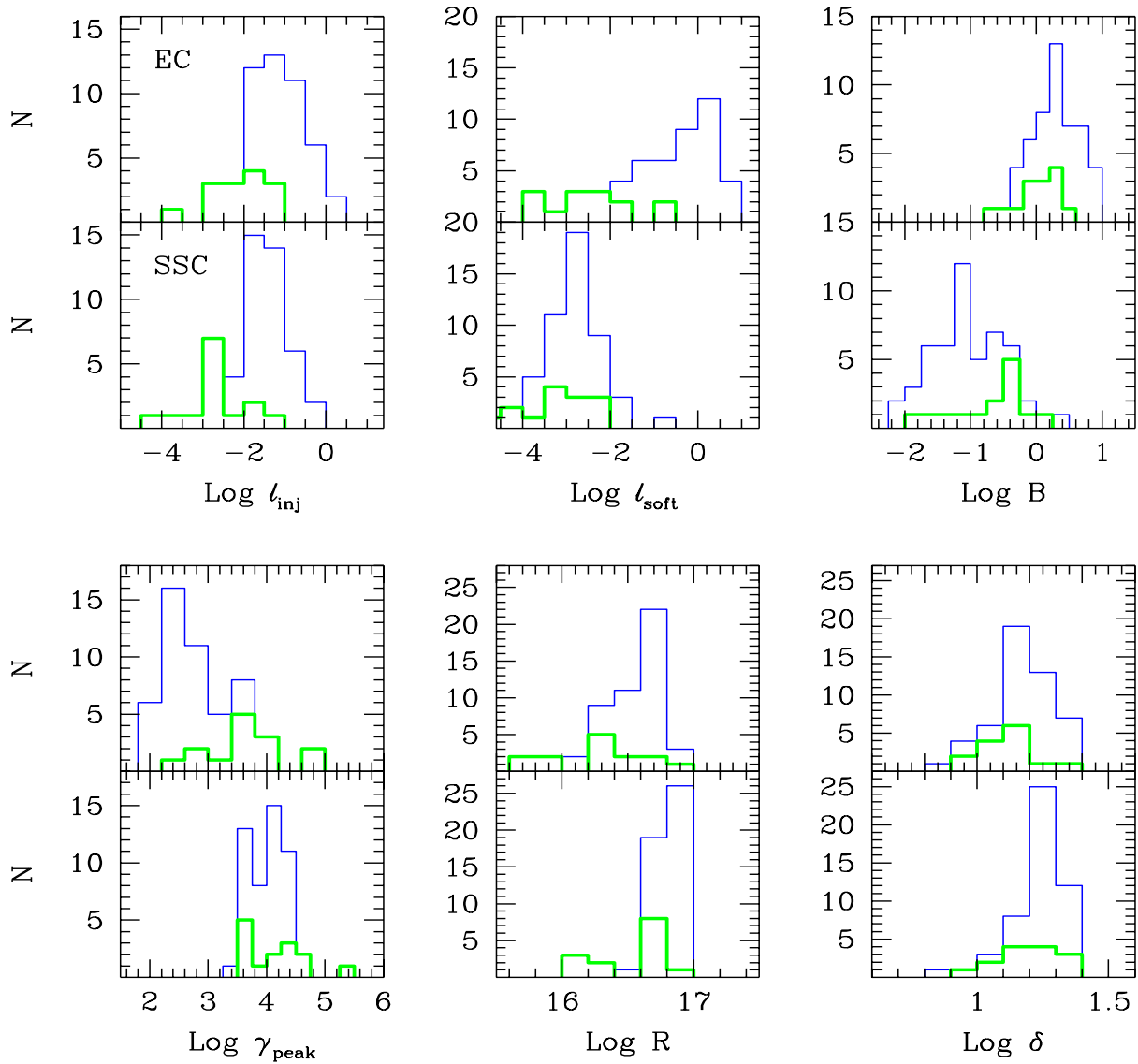}
\caption[h]{The histograms show the distributions of the parameters
of the fits for the EC (upper panels) and the SSC (lower panels)
models for all sources.  Thick lines represent BL Lac objects.
$\ell_{\rm soft}$ is $\ell_{\rm ext}$ in the case of the EC model
and $\ell_{\rm syn}$ (compactness of the synchrotron radiation)
in the case of the SSC model}
\end{figure*}
 

Summarizing, for the pure SSC model 7 input parameters are required,
namely: $R$, $B$, $\Gamma$, $\ell_{\rm inj}$, $s$, $\gamma_{\rm min}$,
$\gamma_{\rm max}$.  If the inverse Compton scattering on external
photons (EC model) is included, two more parameters are required
$\ell_{\rm ext}$ and $x_{\rm ext}$ (i.e. a total of 9).

However, if the slope of the injected electron distribution is steep
($s>2$), which is the case for most sources, the exact value of
$\gamma_{\rm max}$ becomes energetically unimportant and is
practically irrelevant in the comparison with spectral data.
Furthermore, if the external radiation is constituted of broad line
photons, the value of $x_{\rm ext}$ is constrained in a very narrow
energy range.

In conclusion, even if there are formally nine free input parameters,
$\gamma_{\rm max}$ is relatively unimportant, $x_{\rm ext}$ is tightly
limited and constraints apply to the possible values of $R$, $\delta$
and $\ell_{\rm inj}$.

Given the number of free parameters, a key question concerns the
uniqueness of the `fits'. 
 
From an observational point of view, we have already mentioned the
most critical quantities, namely the energy and luminosity of the two
spectral peaks, which determine the global spectral shape.
Furthermore, the optical--to--X--rays and the
X--ray--to--$\gamma$--ray spectral indices and the three limits
discussed above (on $R$, $\delta$, $\ell_{\rm inj}$) also
constrain the parameters.

In particular, as shown by Ghisellini et al. (1996), in the SSC
scenario all the parameters are strongly constrained by the
frequencies of the synchrotron and the self--Compton peaks, and by the
corresponding powers.  These allow to uniquely determine $B$, $\delta$
and $\ell_{\rm inj}$.  $R$ and $\ell_{\rm inj}$ have then to satisfy
the above constraints.

In the EC scenario, the further free parameter $\ell_{\rm ext}$ 
could be in principle be constrained by the observed
soft photon component (e.g. emission line intensities).  Although 
we choose not to assume a priori the origin of the external
soft photon field, however this should at least contain the
contribution of photons produced in the broad line region.  The
addition of $\ell_{\rm ext}$ as a free parameter makes the choice of
$\delta$ not unique.

Independently of our assumption of selecting values of $\delta$
consistent with the observed superluminal speeds, we have therefore
examined the consequences of allowing arbitrary values of $\delta$.
In principle, one can obtain good fits with, e.g., $\delta\sim 100$
and small values of $\ell_{\rm inj}$ ($\propto \delta^{-4}$).  
But since $\ell_{\rm ext}\propto\Gamma^2\sim \delta^2$, the external
photons become important as targets in collisions with very high
energy $\gamma$--rays, with the result of overproducing X--rays
(see \S 3.4 and Ghisellini \& Madau 1996).
Assuming instead a small value of $\ell_{\rm ext}$ corresponds to
limit the broad line radiation to an implausibly small contribution.
We therefore conclude that also the EC model is well constrained.

The main source of uncertainty on the model parameters is given by the
incomplete and poor spectral coverage of several sources, which does
not allow to determine with accuracy the observational constraints,
most critically the peak of the synchrotron emission.  Another source
of uncertainty regards the presence of other spectral components.  Our
single--zone and homogeneous models cannot reproduce the spectrum at
frequencies below the far IR, since at these frequencies the model
spectrum is self--absorbed.  Other emitting regions, of greater
dimensions, are necessary to fit the radio band.  If these also emit
in the IR and optical bands, they could contribute to the synchrotron
spectrum, possibly shifting its peak at a frequency different from the
one corresponding to the $\gamma$--ray emitting region.

As far as the actual fitting procedure is concerned, it should be
pointed out that we try to reproduce collections of data that rarely
are simultaneous, for sources whose variability is a defining
property.  Note also that the spectral coverage differs widely from
object to object.  On one hand this makes it difficult to determine a
method to estimate the goodness of `fit' other than the visual inspection
(with all its limits).  
On the other hand, as mentioned in the
introduction, the main goal of our study is not to model specific
blazars, but to unveil possible trends using a large number of
sources.  In this sense, although the model parameters relative to a
specific source are not `objectively' found (as it would be for a best
fit determined through a statistical $\chi^2$ test) nevertheless they
represent a reasonable description of the spectral properties of each
of the blazar in the sample.

\begin{figure*} 
\psfig{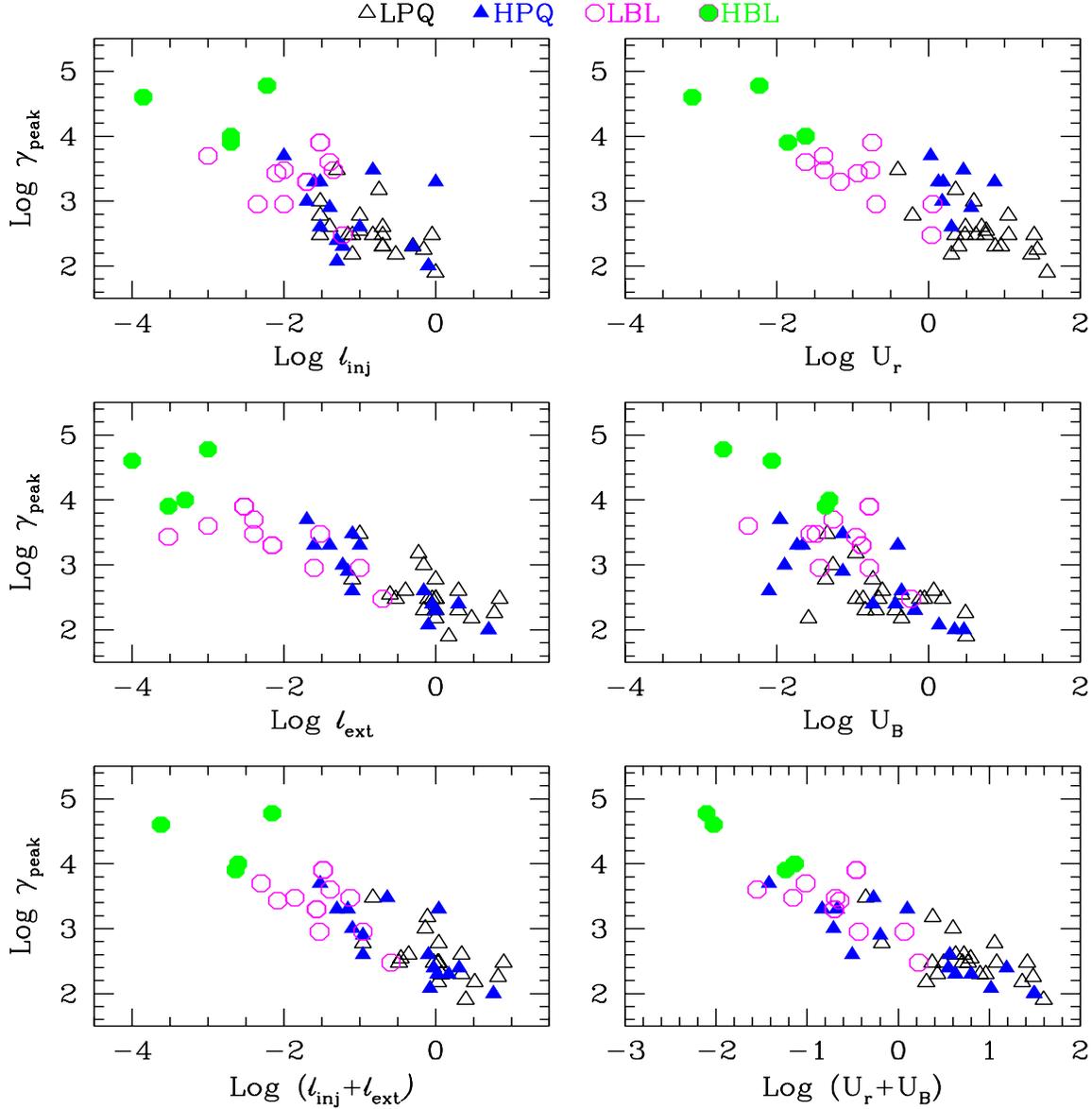}
\caption[h]{$\gamma_{\rm peak}$ (from the EC model) is plotted against
some other, intrinsic, parameters.  The statistical significance of the
correlations is reported in Table~3, in the Appendix.}
\end{figure*}

\section{Results}

The spectral distributions derived from the SSC and EC models which
better describe the SEDs are superposed to the data in Fig.~1a-f.  On
the basis of these `fits' $both$ models can be acceptable for
basically all sources.
As mentioned in the Introduction, the EC model includes the
contribution of the SSC component, i.e. the inverse Compton flux
is calculated assuming, as seed photon flux, the sum of the internal
synchrotron photons and the externally produced photons.
As a rule, the EC component is always dominating at the highest 
frequencies ($\gamma$--ray band), and often it also dominates
in the X--ray band. 
Note however that in some sources, as 
0208--512, 0420--014, 0440--003, 0735+178, 0954+658,
1156+295, 1253--055 and 2032+107, the X--ray band is mainly produced
by the SSC component even in the EC model.

In Table~2 (in the Appendix) the input parameters for all the fits are
reported.  The most remarkable difference between the two sets of
parameters (SSC vs EC) is the relatively smaller value of the magnetic
field in the SSC model.  This has to be expected, since in order to
reproduce the large ratios of inverse Compton to synchrotron
luminosities, the SSC model requires a small magnetic energy density,
while this constraint is relaxed in the EC scenario.

Although at first sight it seems difficult to discriminate between the
two models, at least for FSRQ the parameters derived in the SSC
scenario argue in favour of the EC model.  In fact, let us consider
the typical quantities required by the SSC model: $\delta\sim 20$,
$\ell_{\rm inj} \sim 0.03$, $B\sim 0.05$ G, $R\sim 10^{17}$ cm.  These
imply a compactness in synchrotron radiation $\ell_{\rm syn} \sim
3\times 10^{-3}$.  Consequently, inverse Compton scattering on broad
line photons is unimportant if the external radiation energy density
$U_{\rm ext}$ (as seen in the comoving frame) is less than the
synchrotron one, i.e.:
$$
L_{\rm BLR}\, < \, {m_{\rm e} c^3 \ell_{\rm syn} \over \sigma_{\rm T}
R \Gamma^2}\, R_{\rm BLR}^2 \simeq 3\times 10^{42}\, R_{\rm BLR,18}^2
\quad {\rm erg\, s^{-1}}, 
\eqno(5)
$$
where the above typical parameters have been used and $R_{\rm BLR}=
10^{18}R_{\rm BLR,18}$ cm.  This limit on $L_{\rm BLR}$ is certainly
not observationally satisfied in FSRQ (e.g. Celotti, Padovani \&
Ghisellini 1997).  For BL Lacs the situation is ambiguous.  While the
absence of observable emission lines in most BL Lacs suggests that the
SSC process can dominate on the EC one, (weak) broad emission lines
have been occasionally observed in some LBL (e.g. BL Lac itself,
Vermeulen et al. 1995, Sitko \& Junkkarinen 1985; PKS 0537--441,
Stickel, Fried \& K\"uhr 1993), sometimes exceeding the above limit.
And indeed in some cases the inclusion of an external radiation
component yields a better broad band fit.  Therefore in the following
the discussion is focused on the results of the EC scenario.
or HBL the different parameters derived in the SSC and EC models can
be considered an indication of the allowed range of values and, in
particular, the external radiation (e.g emission line luminosity)
required by the EC fit can be taken as an upper limit.

Fig.~2 shows the distributions of values of the model parameters
within the EC (upper panels) and the SSC (lower panels) scenarios.
The thick solid lines correspond to the results for BL Lacs.  In the
EC case, BL Lac objects almost always form the left tail of the
distributions, being characterized by smaller compactness, magnetic
field and slightly smaller degree of beaming.  On the contrary,
comparable dimensions $R$ and greater value of $\gamma_{\rm peak}$ are
required by the EC fits of BL Lacs with respect to FSRQ.  In the SSC
case the required $\gamma_{\rm peak}$ is limited in a narrow range,
without a clear distinction between BL Lacs and FSRQ, while BL Lacs are
characterized by a larger average value of the magnetic field.

As a consequence of the constraints imposed on the Doppler factor and
the variability timescales, the distributions of $\delta$ and $R$ span
less than one order of magnitude each, with $R\sim 10^{16-17}$ cm
\footnote{Note that given the high values of the Doppler factors
derived from the fits, the assumption $\theta \sim 1/\Gamma$ is
satisfied. The only exception is 0521--365, which only requires
$\delta\sim 1.4$}.  On the contrary the other (intrinsic) quantities
are spread over much larger ranges of values, with the external photon
compactness covering the wider interval of about 5 decades.

The injected particle energy distribution is highly different from
source to sources (see Table 2), in shape, compactness and (rather
low) maximum energy $\gamma_{\rm max} m_{\rm e} c^2$, thus not
requiring a very finely tuned injection/acceleration mechanism.

\subsection{Correlations}

The main goal of this work is to determine trends and correlations
among physical quantities which can shed light on the relationship
among different sub--classes of blazars and ultimately on the
processes at work in these objects.

We have found that the most interesting quantity to investigate links
among adopted and derived model parameters is the Lorentz factor at
the break of the electron distribution $\gamma_{\rm peak}$, which
determines the location of both the synchrotron and the Compton peaks,
and therefore largely determines the shape of the SED.  The other
important parameters controlling the SED are the ratio of the Compton
to synchrotron powers, i.e. the Compton dominance $L_{\rm C}/L_{\rm
syn}$, the power (or the corresponding compactnesses) injected in
the form of electrons (which in our model corresponds to the radiated
power), and the power in the external photon component.  The results
of linear correlations involving these quantities are shown in
Figs.~3 and 4 and their statistical significance is reported in
Table~3 in the Appendix.  For completeness, we list in Table~4 (in the
same Appendix) also the results of the linear correlations in the case
of the SSC model.

\begin{figure*} 
\psfig{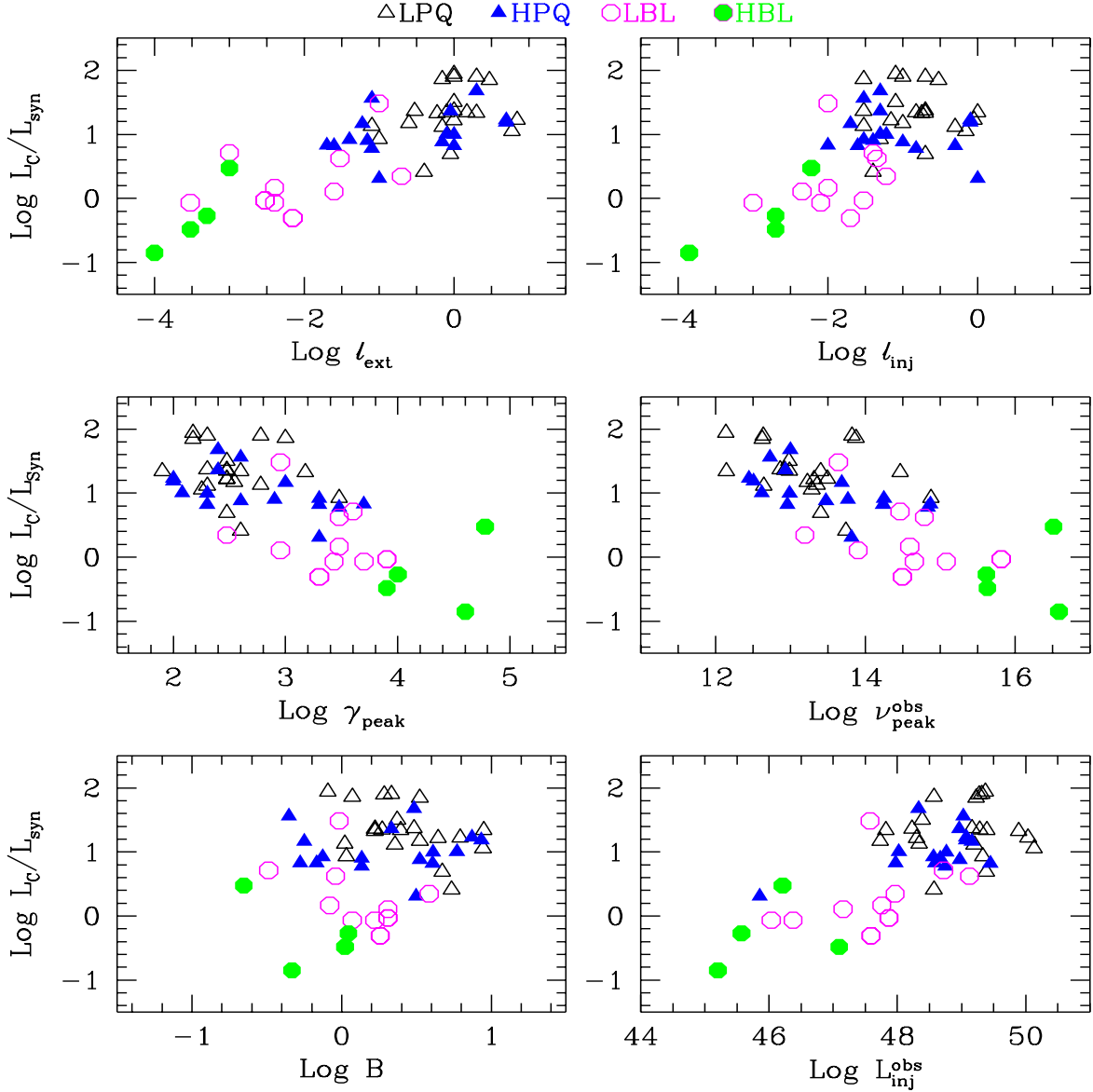}
\caption[h]{The Compton dominance $L_{\rm C}/L_{\rm syn}$ as derived
from the EC model is plotted against some other (both intrinsic and
`observable') parameters.  The statistical significance of the
correlations is given in Table~3.}
\end{figure*}

Let us consider the results of the correlations: 

\begin{enumerate}

\item $\gamma_{\rm peak}$ --- Strong correlations are present between
$\gamma_{\rm peak}$ and $\ell_{\rm ext}$, $\ell_{\rm inj}$ and the
energy densities in radiation $U_{\rm r}$ and magnetic field $U_{\rm
B}$, both for the whole sample and for the FSRQ sub--sample.
In particular, a very strong linear correlation is found between
$\gamma_{\rm peak}$ and the total energy density, with a dependence
$\gamma_{\rm peak}\propto (U_{\rm r}+U_{\rm B})^{-0.6}$.  The same
trend appears from the correlation of $\gamma_{\rm peak}$ with
$\ell_{\rm ext}+\ell_{\rm inj}$.  It should be pointed out that these
correlations are not, or at most only partly, induced by an
observational selection effect: there would not be bias against
detecting sources with either high values of $\gamma_{\rm peak}$ and
$(U_{\rm r}+U_{\rm B})$ or viceversa.  Furthermore the significance of
the correlations (i.e. their small spread) can be taken as a
posteriori indication of the tightness of the observational
constraints imposed on the model parameters.  Note that HBL, LBL and
quasars are located along a sequence.

\item $L_{\rm C}/L_{\rm syn}$ --- The Compton dominance correlates
with $\gamma_{\rm peak}$, $\ell_{\rm ext}$, $\ell_{\rm inj}$, and
$\nu_{\rm peak}^{\rm obs}$, the latter being the $observed$ peak
frequency of the modeled synchrotron emission.  It also correlates
with the observed (beamed) power $L_{\rm inj}^{\rm obs} = L_{\rm inj}
\delta^4$, while only a weak correlation exists between $L_{\rm
C}/L_{\rm syn}$ and the magnetic field intensity.  The statistical
significance of all these correlations is higher when considering the
entire blazar sample, while weakens when the subsamples of BL Lacs and
FSRQ are considered separately.  Again, note that in all cases BL Lacs
are `separated' from FSRQ, with HBL at the extremes and some LBL
smoothly overlapping with FSRQ.

\item $\ell_{\rm ext}$ vs $\ell_{\rm inj}$ --- A significant linear
correlation is present when FSRQ are considered, while (most) BL Lacs
show a relative deficiency in the external photon component with
respect to this trend (see Fig.~5).

\end{enumerate}



To further investigate the correlations among the various quantities
described above, we ran a principal component analysis (PCA) program
on the correlation matrix.
Briefly, the PCA is a method to describe a multidimensional ensemble
of correlated parameters, by defining a new coordinate system in which
each successive coordinate direction defined by the eigenvectors,
explains as much of the remaining variance in the data is possible.
PCA reduces the number of relevant components and the remaining should
represent more basic parameters than the original ones. (see e.g.
Boroson and Green 1992 for an application).
We choose to present the PCA done with the 6 most important parameters
of the fits:
$\ell_{inj}$, $\ell_{ext}$, $\delta$, $\gamma_{peak}$, $B$ and $R$.

The results of the analysis are presented in Table 6 of the
Appendix which lists the
most significant eigenvectors in terms of their projection
upon the original 6 quantities.
At the top of each column the percentage variance accounted for by
the eigenvectors is given.

The first eigenvector accounts for about 45\% of the total variance
and is dominated by the contribution of the two compactnesses and the
magnetic field energy density which anti--correlate (see above) with
$\gamma_{\rm peak}$.
This eigenvector could be associated with the total power of the source.
The largest contribution to the second eigenvector comes from $\delta$,
$R$ and $\gamma_{peak}$, while the only relevant projection on
the third one is due to $\gamma_{\rm peak}$.

Given that the Doppler factor $\delta$ and the blob dimension $R$
are not completely independent quantities (see section 3.4), the PCA
analysis points to $\gamma_{\rm peak}$, $\ell_{\rm inj}$ and $\ell_{\rm ext}$
as fundamental variables in explaining the formation of the blazar SED,
and confirms the results found through the linear regression analysis.

\section{Discussion and conclusions}
\subsection{Generalities}

We examined and reproduced the broad band spectral properties of a
sample of 51 $\gamma$--ray loud blazars, in terms of synchrotron and
inverse Compton emission from a homogeneous (one--zone) model.  This
gives a reasonable good description of the observations at frequencies
greater than typically $\sim 10^{11}$ Hz.  The radio emission is
expected to be produced by less compact regions on larger jet scales.

On the basis of the `fits', it is difficult to determine whether the
seed field for the inverse Compton scattering is mostly provided by
synchrotron or external photons.  However, as pointed out by Sikora et
al. (1994), the presence of broad emission lines, together with the
found values of $\delta$, make photons produced externally to the
emitting jet unavoidably important.  For this reason we consider the
parameters derived in the EC scenario as the most likely for all
quasars and some BL Lacs.

Strong correlations have been found among the physical parameters
derived from the EC model.

\begin{figure}
\psfig{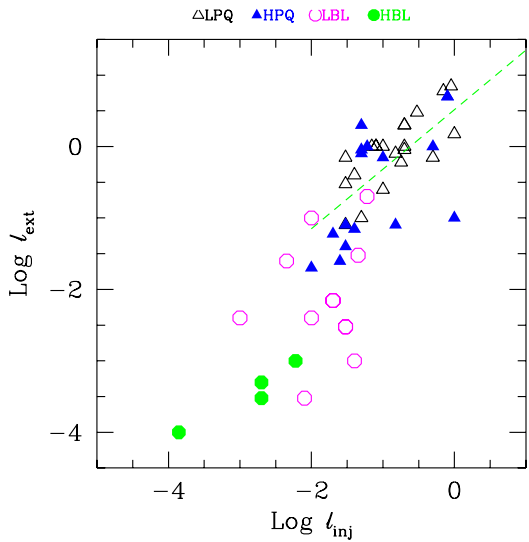}
\vskip -1 truecm
\caption[h]{The relation between the compactnesses in external photons
$\ell_{\rm ext}$ and injected power $\ell_{\rm inj}$. The line
represents the best--fit linear correlation for FSRQ only.}
\end{figure}

Of particular physical interest is the strong correlation between the
energy of electrons emitting at the peak of the observed spectra and
the total energy density present in the emitting region, $\gamma_{\rm
peak}\propto (U_{\rm r}+U_{\rm B})^{-0.6}$. One way to explain this is
to assume that $\gamma_{\rm peak}$ is the result of a competition
between the radiative cooling and the (re--)acceleration process, i.e.
$\dot\gamma_{\rm acc}(\gamma_{\rm peak}) \sim \dot\gamma_{\rm
cool}(\gamma_{\rm peak})$.  The typical emitting electron would in
this case be quickly accelerated up to the energy where cooling is
important, while only a few particles would be accelerated at higher
energies.  The found correlation would then imply that the
(re--)acceleration process is almost independent of both the energy
density in the region (both in radiation and in magnetic field) and
the energy of the particles, since $\dot\gamma_{\rm cool}(\gamma_{\rm
peak})\propto \gamma^2_{\rm peak}(U_{\rm r}+U_{\rm B}) \sim$ const.
In addition, the injected particle distribution does not require
characteristic shape and/or maximum energy.  We postpone the discussion
of this interesting result to future work; here we would only like to
mention the `hot jet' model put forward by Sikora et al. (1997), in
which the balance between heating and cooling can lead to a formation
of a peak in the electron energy distribution.

We found another strong correlation between $\gamma_{\rm peak}$ and
the Compton dominance $L_{\rm C}/L_{\rm syn}$.  On one side this
simply confirms and quantifies, from a different perspective, the
observational trends pointed out by e.g. Fossati et al. (1998) on the
relation between the dominance of the Compton/$\gamma$--ray emission
and the energy of the peaks of the two spectral components for
complete samples of blazars.  On the other side, it suggests that this
link can be simply interpreted as the consequence of a change in the
radiation energy density of the external field.  An increase in the
latter in fact leads to an increase in the particle Compton cooling
and therefore both to a decrease in $\gamma_{\rm peak}$ and a relative
increase in the $\gamma$--ray luminosity.  Once again we stress, as
discussed in the next Section, that different sub--classes of blazars
are located in different areas of this correlation.

As expected from the above correlations $\gamma_{\rm peak}$ is also
(inversely) related to the power injected in the form of relativistic
emitting particles.

As presented in Fossati et al. (1998), the ratio of the frequency of
the Compton ($\nu_{\rm C}$) to the synchrotron ($\nu_{\rm syn}$) peak
is compatible with being approximately constant.
Our results are in agreement
with these findings, despite the relatively wide range spun by
$\gamma_{\rm peak}$ ($\sim$ 3 decades for the EC model).
In fact in the EC scenario (and in the Thomson regime of the inverse
Compton process), 
$\nu_{\rm C}/\nu_{\rm syn}\sim \Gamma \nu_{\rm ext}/\nu_{\rm B}$, 
is independent of $\gamma_{\rm peak}$ (here
$\nu_{\rm B}=eB/(2\pi m_{\rm e}c)$ is the cyclotron frequency),
and the narrow range of values of $B$ found in the EC model can
account for the approximate constant ratio of 
$\nu_{\rm C}/\nu_{\rm syn}$.

In the SSC model, instead, it is 
$\gamma_{\rm peak}$ which is found in a narrow range 
(less than 2 decades).
In the SSC case, we expect 
$\nu_{\rm C}/\nu_{\rm syn}\sim \gamma^2_{\rm peak}$ 
(in the Thomson regime), and
$\nu_{\rm C}/\nu_{\rm syn}\sim \gamma^{-1}_{\rm peak}$ 
in the extreme Klein--Nishina regime (in this case 
$h\nu_{\rm C}\sim \gamma_{\rm peak} m_{\rm e} c^2$).

\subsection{The blazar unification}

Evidence for continuity in the observed spectral properties of BL Lacs
and FSRQ have been recently found by Fossati et al. (1998), by
studying complete samples of sub--classes of blazars in different
energy bands. Diagrams and quantities derived by Fossati et
al. (1998) from either data or their analytical representation turn
out to be similar to and consistent with those found in this paper
through model fitting. In Fig.~6, the model considered here is applied
to the average SEDs derived by Fossati et al. (1998) by binning,
according to the radio luminosity, both BL Lacs and FSRQ belonging to
complete samples. The parameters of these fits are reported in Table~5
in the Appendix. The fact that the model fits the average SEDs
derived from complete blazars samples, with similar parameters and
trends as for the $\gamma$--ray blazars, gives us confidence
that our results are valid for all blazars.

The main result of the present paper concerns the intrinsic
relationship among phenomenologically different classes of blazars,
and in particular the evidence for a well defined sequence in the
properties of HBL, LBL and FSRQ with increasing importance of an
external radiation field: the observed spectral properties of HBL,
LBL, HPQ and LPQ can be therefore accounted for by e.g. the increasing
role of broad emission line radiation (see also Fig.~6 and Table~5).
This in fact dictates the peak energy of the emitting particle
distribution and hence the shape of the spectra, thus determining the
classification of an object into one of the blazar flavors.  The
fundamental physical processes occurring in and outside the
relativistic jet are instead the same.  This is indicated by the
correlation between $\ell_{\rm ext}$ and $\ell_{\rm inj}$, which seems
to `set in' for more powerful object, from LBL up to the most luminous
LPQ (see Fig.~5).

This proposed blazar unifying sequence can be therefore summarized as
follows (see the schematic sketch in Fig.~7):


\begin{figure}
\psfig{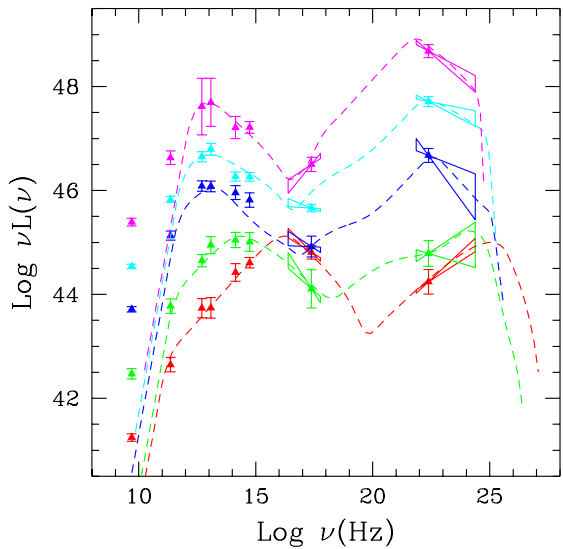}
\caption[h]{Fits with the EC model of the average SEDs derived by
Fossati et al. (1998). BL Lacs and FSRQ belonging to complete samples
have been divided in bins accordingly only to their radio luminosity,
and the average fluxes in each bin have been computed. The model
parameters are reported in Table~5 (Appendix), and are in complete
agreement with the parameters determined for the $\gamma$--ray loud
sources examined in this work.}
\end{figure}

\begin{enumerate}

\item HBL are sources characterized by the lowest intrinsic power and
the weakest external radiation field (no or weak emission lines).
Consequently the cooling is less dramatic and particles can be present
with energies high enough to produce synchrotron emission extending to
soft X--ray energies and TeV radiation through the SSC process. Being
the inverse Compton cooling ineffective, the Compton dominance is
expected to be small;

\item LBL are intrinsically more powerful than HBL and in some cases
the external field can be responsible for most of the cooling.  The
stronger cooling limits the particle energy implying that the
synchrotron and inverse Compton emission peak at lower frequencies, in
the optical and GeV bands, respectively, with a larger Compton
dominance parameter;

\item FSRQ represent the most powerful blazars, where the contribution
from the external radiation to the cooling is the greatest.  The
emission by synchrotron and EC cannot extend at frequencies larger
than the IR and MeV--GeV bands and the $\gamma$--ray radiation
completely dominates the radiative output.  Within this class, there
is an hint of a further subdivision between low and high polarization
objects, with a tendency for LPQ to be more extreme (lower values of
$\gamma_{\rm peak}$ and larger values of $(U_{\rm r}+U_{\rm B})$,
$\ell_{\rm inj}$ and so on).

\end{enumerate}

\begin{figure}
\psfig{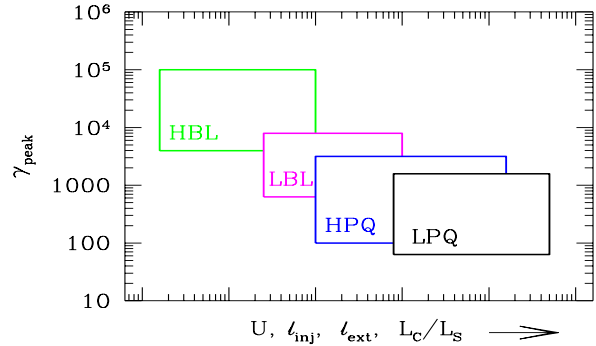}
\vskip -2 true cm
\caption[h]{Schematic representation of the proposed unifying scheme:
the sequence HBL, LBL, HPQ, LPQ corresponds to an increase in the
external radiation field, the total energy density and the injected
power. These in turn result in a decrease of $\gamma_{\rm peak}$ and
an increase in the Compton dominance.}
\end{figure}

The correlations among the different quantities ensure that the
knowledge of one of them allows to estimate the entire spectral energy
distribution, and also the probable classification of the object.
This is of course of great relevance for the study at high energies of
those blazars not detected so far in the $\gamma$--ray band and the
consequences on their variability patterns and duty cycles.  Finally,
the above findings have to be taken into account when considering the
absorption of high energy radiation by the diffuse background fields
as well as the estimates on the blazar contribution to the
$\gamma$--ray background.

In the currently most accepted unification schemes for radio--loud
sources, weak and powerful blazars are the beamed counterparts of
Fanaroff--Riley type I (FR~I, Fanaroff \& Riley 1974) and type II
(FR~II) radio galaxies, respectively. Indeed continuity in the
properties of blazars along a power sequence has been suggested by
Maraschi \& Rovetti (1994), Sambruna et al. (1996) and Fossati et
al. (1997) on the basis of statistical arguments.  Within this frame,
the blazar sequence would therefore manifest itself in several
observational properties, including the total source power, the
luminosity in emission lines, the extended radio power, the dominance
of $\gamma$--rays over the other spectral components and the broad
band shape of the SED (see Fig.~7).  We therefore provide evidence for
the unification of all radio--loud sources and suggest a deeper
physical understanding for it, based on the total power generated in
the very central engine of these spectacular sources.

\section*{Acknowledgments}
Thanks are due to the Italian MURST (Annalisa Celotti, GF) and the
Institute of Astronomy PPARC Theory Grant (Annalisa Celotti) for
financial support.  
Andrea Comastri acknowledges financial support from the
Italian Space Agency under contract ARS-96-70
This research has made use of the NASA/IPAC
extragalactic database (NED), which is operated by the Jet Propulsion
Laboratory, California Institute of Technology, under contract with
the National Aeronautic and Space Administration.

\section*{References}

\refitem Adam G., 1985, A\&AS, 61, 225 (A85)

\refitem Aller H.D, Aller M.F., Latimer G.E., Hodge, P.E., 1985, ApJS,
59, 513 (Al85)

\refitem Bersanelli M., Bouchet P., Falomo R., Tanzi E.G., 1992, AJ, 104, 28
(Be92)

\refitem Bertsch D.L. et al., 1993, ApJ, 405, L21 (Be93)

\refitem Biermann P.L., Kuhr H., Snyder W.A., Zensus J.A., 1987, A\&A, 85, 9 (Bi87)

\refitem Blandford R.D., 1993, in Friedlander M., Gehrels N., Macomb D.J., eds,
Proc. CGRO AIP 280. New York, p. 533

\refitem Blandford R.D., Levinson A., 1995, ApJ, 441, 79

\refitem Bloom S.D., Marscher A.P., 1991, ApJ, 366, 16 (BM91)

\refitem Bloom S.D., Marscher A.P., 1993, in Friedlander M., Gehrels
N., Macomb D.J., eds, Proc. CGRO AIP 280. New York, p. 578

\refitem Bloom S.D., Marscher A.P., Gear W.K., Terasranta H., Valtaoja E.,
Aller H.D., Aller M.F., 1994, AJ, 108, 398 (BM94)

\refitem Boroson T.A. \& Green R., 1992, ApJS, 80, 109

\refitem Bozyan E.P., Hemenway P.D., Argue A.N., 1990, AJ, 99, 1421 (Bo90)

\refitem Bregman J.N., Glassgold A.E., Huggins P.J., Kinney A.L., 1985,
      ApJ, 291, 505 (B85)

\refitem Breslin A.C. et al., 1997, IAUC 6592 (B97)

\refitem Brinkmann W., Siebert J., Boller T., 1994, A\&A, 281, 355 (B94)

\refitem Brinkmann W., Siebert J., Reich W., Furst E., Reich P., Voges W.,
     Trumper J., Wielebinski R., 1995, A\&AS, 109, 147 (B95)

\refitem Cappi M., Comastri A., Molendi S., Palumbo G.C.C., Della
Ceca R., Maccacaro T., 1994, MNRAS, 271, 438

\refitem Catanese M. et al., 1997a, ApJ, 480, 562 (Ca97)

\refitem Catanese M. et al., 1997b, proceedings of 25th ICRC (Durban), 
in press (Cat97)


\refitem Celotti A., Padovani P., Ghisellini G., 1997, MNRAS, 286, 415

\refitem Chini R., Biermann P.L., Gemund H.-P., 1989, A\&A, 221, L3 (Ch89)

\refitem Clegg P.E. et al., 1983, ApJ, 273, 58 (C83)

\refitem Collmar, W., 1996, in Gamma ray emitting AGN, MPI H -V37-1996
eds. J.G. Kirk, M. Camenzind, C. von Montigny \& S. Wagner, p. 9 (Co97)

\refitem Comastri A., Fossati G., Ghisellini G., Molendi S., 1997, ApJ, 480, 
534 (C97) 

\refitem Condon J.J., Hicks P.D., Jauncey D.L., 1977, AJ, 82, 692 (Co77)

\refitem Condon J.L., Anderson E., Broderick J.J., 1995, AJ, 109, 2318 (Co95)

\refitem Dermer C.D., 1995, ApJ, 446, L63

\refitem Dermer C.D., Schlickeiser R., 1993, ApJ, 416, 458

\refitem Dingus B.L. et al., 1996, ApJ, 467, 589 (Di96)

\refitem Dondi L. \& Ghisellini G., 1995, MNRAS, 273, 583

\refitem Edelson R.A., 1994, AJ, 94, 1150 (E94)

\refitem Elvis M., Plummer D., Schachter J., Fabbiano G., 
1992, ApJS, 79, 331 (E92)

\refitem Falomo R., Scarpa R., Bersanelli M., 1994, ApJS, 93, 125 (F94)

\refitem Fanaroff B.L., Riley J.M., 1974, MNRAS, 167, 31 


\refitem Fichtel C.E. et al., 1994, ApJS, 94, 551

\refitem Fossati G., Maraschi L., Celotti A., Comastri A., Ghisellini G., 
1998, MNRAS, in press

\refitem Fossati G., Celotti A., Ghisellini G. \& Maraschi, L., 
1997, MNRAS, 289, 136
                                                    
\refitem Friecke K.J., Kollatschny W.,  Witzel A., 1983, A\&A, 117, 60 (Fr83)

\refitem Giommi P., Ansari S.G., Micol A., 1995, A\&AS, 109, 267

\refitem Gear W.K. et al., 1994, MNRAS, 267, 167 (G94)

\refitem Ghisellini G., 1989, MNRAS, 238, 449

\refitem Ghisellini G., Madau P., 1996, MNRAS, 280, 67

\refitem Ghisellini G., Maraschi L., 1989, ApJ, 340, 181

\refitem Ghisellini G., Maraschi L., 1994, in  Fichtel C.E., 
Gehrels N.,  Norris J.P., eds, The Second Compton
Symp., AIP 304. New York, p. 616

\refitem Ghisellini G., Maraschi L., Tanzi E., Treves A., 1986, 
ApJ, 310, 317  (G86)

\refitem Ghisellini G., Maraschi L., Dondi L., 1996, A\&AS, 120, 503

\refitem Ghisellini G. et al., 1997, A\&A, in press (G97)


\refitem Glass I.S., 1981, MNRAS, 194, 795 (G81)

\refitem Hartmann R.C. et al. 1993, ApJ, 407, L41 (Ha93)

\refitem Hunter S.D. et al., 1993, A\&A, 272, 59 (H93)

\refitem Impey C.D., Neugebauer G., 1988, AJ, 95, 307 (IN88)

\refitem Impey C.D., Tapia S., 1988, ApJ, 333, 666 (IT88)

\refitem Impey C.D., Tapia S., 1990, ApJ, 354, 124 (IT90)

\refitem K\"uhr H., Witzel A., Pauliny-Toth I.I.K., Nauber U., 1981,
      A\&AS, 45, 367 (K81)

\refitem Landau R.,  Jones T.W., Epstein E.E., Neugebauer G., Soifer B.T.,
Werner M.W., Puschell J.J., Balonek T.J., 1983, ApJ, 268, 68 (L83)

\refitem Landau R. et al., 1986, ApJ, 308, 78 (L86)

\refitem Lawrence A., Rowan-Robinson M., Efstathiou A., Ward M.J., Elvis M.,
Smith M.G., Duncan W.D., Robson E.I., 1991, MNRAS, 248, 91 (L91)

\refitem Lawson A.J., Turner M.J.L., 1997, MNRAS, 288, 920 (LT97)

\refitem Ledden J.F., O'Dell S.L., 1985, ApJ, 298, 630 (Le85)

\refitem Lin Y.C. et al., 1995, ApJ, 442, 96 (L95)

\refitem Lin Y.C. et al., 1996, ApJS, 105, 331 (L96)

\refitem Litchfield S.J., Robson E.I., Stevens J.A., 1994, MNRAS, 270, 341 
(Li94)

\refitem Lorenzetti D., Massaro E., Perola  G.C., Spinoglio L., 1990,
A\&A, 235, 35 (L90)

\refitem McNaron-Brown K. et al., 1995, ApJ, 451, 575 (Mc95)

\refitem Macomb D.J. et al., 1995, ApJ, 449, L99 (Ma95)

\refitem Macomb D.J. et al., 1996, ApJ, 459, L111 (Erratum) (Ma96b)

\refitem Madejski G., Takahashi T., Tashiro M., Kubo H., Hartman R.,
      Kallman T., Sikora M., 1996, ApJ, 459, 156 (Ma96)

\refitem Mannheim K., 1993, A\&A, 269, 67

\refitem Maraschi L., Rovetti F., 1994, ApJ, 436, 79

\refitem Maraschi L., Schwartz D.A., Tanzi E.G., Treves A.,
1985, ApJ, 294, 615 (Ma85)

\refitem Maraschi L., Ghisellini G.,  Celotti A., 1992, ApJ, 397, L5

\refitem Maraschi L. et al. 1994, ApJ, 435, L91 (Ma94)

\refitem   Maraschi L., Fossati G., Tagliaferri G., Treves A., 
1995, ApJ, 443, 578

\refitem Mattox J.R. et al., 1993, ApJ, 410, 609 (M93)

\refitem Mattox J.R, Wagner S.J., Malkna M., McGlynn T.A., Schachter J.F.,
Grove J.E., Johnson W.N., Kurfess J.D., 1997, ApJ, 476, 692 (M97)

\refitem Mattox J.R., Schachter J., Molnar L., Hartman R.C., Patnaik
A.R., 1997, ApJ, 481, 95

\refitem Mukherjee R. et al., 1995, ApJ, 445, 189 (Mu95)

\refitem Mukherjee R. et al., 1996, ApJ, 470, 831 (Mu96)

\refitem Nolan P.L. et al., 1993, ApJ, 414, 82 (N93)

\refitem Nolan P.L. et al., 1996, ApJ, 459, 100 (N96)

\refitem Netzer H. et al., 1996, MNRAS, 279, 429 (Ne96)

\refitem Padovani P., Giommi P., 1995, ApJ, 444, 567 

\refitem Perley R.A., 1982, AJ, 87, 859 (P82)

\refitem Perlman E.S. et al., 1996, ApJS, 104, 251 (P96)

\refitem Petry D. et al., 1996, A\&A, 311, L13

\refitem Pian E. et al., 1993, ApJ, 416, 130 (Pi93)

\refitem Pian E., Falomo R., Scarpa R., Treves A., 1994, ApJ, 37, 152 (Pi94)

\refitem Pian E., Falomo R., Ghisellini G., Maraschi L., Sambruna R.M.,
      Scarpa R., Treves A., 1996, ApJ, 459, 169 (Pi96)

\refitem Pica A.J., Smith A.G., Webb J.R., Leacock  R.J., Clements S.,
Gombola P.P., 1988, AJ, 96, 1215 (P88)

\refitem Pohl M., Reich W., Schlickeiser R., Reich P., 
Ungerechts H., 1996, A\&AS, 120, 529 (Po96)

\refitem Quinn J. et al.,  1996, ApJ, 456, L83 (Q96)

\refitem Radecke H.-D. et al., 1995, ApJ, 438, 659 (R95)

\refitem Raiteri C.M. et al., 1997, A\&AS, in press (Ra97)

\refitem Reuter H.-P. et al., 1997, A\&AS, 122, 271 (R97)

\refitem Rieke G.H., Lebofsky M.J., Wi\'sniewski W.Z., 
1982, ApJ, 263, 73 (R82)

\refitem Rybicki G.B. \& Lightman A.P., 1979, Radiative processes in
Astrophysics, J. Wiley \& Sons (New York)

\refitem Robson E.I., Gear W.K., Brown L.M.J., Courvoisier T.J.-L.,
Smith M.G., 1986, Nat, 323, 134 (R86)

\refitem Sambruna R.M., Barr P., Giommi P., Maraschi L.,
      Tagliaferri G.,  Treves A., 1994, ApJS, 95, 371 (Sa94)

\refitem Sambruna R.M. et al., 1997, ApJ, 474, 639 (Sa97)

\refitem Schonfelder V., 1994, ApJS, 92, 593  (Sh94)

\refitem Sikora M., Begelman M.C., Rees M.J., 1994, ApJ, 421, 153

\refitem Sitko M.L., Junkkarinen V.T., 1985, PASP, 97, 1158

\refitem Sitko M.L., Sitko A.K., 1991, PASP, 103, 160 (Si91)

\refitem Smith P.S., Elston R., Berriman G., Allen R.G., Balonek, T.J.,
      1988, ApJ, 326, L39 (Sm88)

\refitem Sreekumar P. et al., 1996, ApJ, 464, 628 (Sr96)

\refitem Stacy J.C., Vestrand W.T., Sreekumar P., Bonnell J., Kubo H.,
       Hartman R.C., 1996, A\&AS, 120, 549 (Sta96)

\refitem Steppe H., Salter C.J., Chini R., Kraysa E., Brunswig W.,
      Lobato Perez J., 1988, A\&AS, 75, 317 (St88)

\refitem Steppe H., Liechti S., Mauersberger R., Kompe C., Brunswig W.,
      Ruiz-Moreno M., 1992, A\&AS, 96, 441 (St92)

\refitem Steppe H. et al., 1993, A\&AS, 102, 611 (St93)

\refitem Stevens J.A., Litchfield S.J., Robson E.I., Hughes D.H., Gear W.K.,
      Terasranta H., Valtaoja E., Tornikoski M., 1994, ApJ, 437, 91 (S94)

\refitem Stickel M., Fried J.W., K\"uhr H., 1993, A\&AS, 98, 393 

\refitem Stickel M., Meisenheimer K., K\"uhr H., 1994, A\&AS, 
105, 211 (St94)

\refitem Stickel M., Rieke G.H., K\"uhr H., Rieke M.J., 1996, ApJ, 468, 556 (St96)

\refitem Terasranta H. et al., 1989, A\&AS, 75, 317 (T89)

\refitem Terasranta H. et al., 1992, A\&AS, 94, 121 (T92)

\refitem Thompson D.J. et al., 1993, ApJ, 415, L13 (T93)

\refitem Thompson D.J. et al., 1995, ApJS, 101, 259 (T95)

\refitem Thompson D.J. et al., 1996, ApJS, 107, 227 (T96)

\refitem Tornikoski M., Valtaoja E., Terasranta H., Lainela M., 
Bramwell D., Botti L.C.L., 1993, AJ, 105, 1680 (To93)

\refitem Tornikoski M. et al., 1996, A\&AS, 116, 157 (To96)

\refitem Tosti G. et al., 1997, A\&AS, submitted (To97)

\refitem Turner M.J.L. et al., 1990, MNRAS, 244, 310 (T90)

\refitem Urry M.C., Padovani P., 1995, PASP, 107, 803

\refitem Urry C.M., Sambruna R.M., Worrall D.M., Kollgaard R.I.,
Feigelson E.D., Perlman E.S., Stocke J.T., 1996, ApJS, 463, 424 (U96)

\refitem Urry C.M. et al. 1997, ApJ, in press (U97) 

\refitem Valtaoja E., Lahteenmaki A., Terasranta H., 1992, 
A\&AS, 95, 73 (V92)

\refitem Vermeulen R.C., Ogle P.M., Tran H.D., Browne I.W.A., Cohen M.H., 
Readhead, A.C.S., Taylor G.B. \& Goodrich R.W., 1995, ApJ, 425, L5

\refitem Vestrand W.T., Stacy J.G., Sreekumar P., 1996, ApJ, 454, L93 (V96)

\refitem Villata M. et al., 1997, A\&AS, 121, 119 (V97)

\refitem von Montigny C. et al., 1995, ApJ, 440, 525 (vM95)

\refitem von Montigny C. et al., 1997, ApJ, in press

\refitem Wagner S., Sanchez-Pons F., Quirrenbach A., Witzel A.,
      1990, A\&A, 235, L1 (W90)

\refitem Wagner S.J. et al., 1995, A\&A 298, 688

\refitem Wagner S.J. et al., 1995, ApJ, 454, L97 (W95)

\refitem Wagner S.J. et al., 1996, AJ, 111, 2187 (W96)

\refitem Wall J.V., Peacock J.A., 1985, MNRAS, 216, 173 (WP85)

\refitem Webb J.R., Smith A.G., Leacock R.J., Fitzgibbons G.L., Gombola P.P.,
Shepherd D.W., 1988, AJ, 95, 374 (W88)

\refitem Webb J.R., Barnello T., Robson I., Hartman R.C., 1995, IAUC 6168
(We95)

\refitem Weekes T.C. et al., 1996, A\&AS, 120, 603

\refitem Wilkes B.J., Tananbaum H., Worral D.M., Avni Y., Oey M.S.,
Flanagan J., 1994, ApJS, 92, 53 (Wi94)

\refitem Wiren S., Valtaoja E., Terasranta H., Kotilainen J.,
     1992, AJ, 104, 111 (W92)

\refitem Worral D.M., Wilkes B.J., 1990, ApJ, 360, 396 (WW90)

\refitem Worral D.M., Giommi P., Tananbaum H., Zamorani G., 1987,
      ApJ, 313, 596 (W87)

\refitem Wright A.E., Wark R.M., Troup E., Otrupcek R., Jennings D.,
      Hunt A., Andcooke D.J., 1991, MNRAS, 251, 330 (W91)

\vskip 1 true cm
\begin{table*}
\centerline{\bf APPENDIX}
\vskip 1 true cm
\caption{List of sources. (1),(2) Source names; (3) redshift; (4)
classification: HPQ and LPQ stand for highly and lowly polarized
quasars, while NP indicates sources with no polarization measure; HBL,
LBL and IBL refer to high, low and intermediate frequency BL Lacs,
respectively; (5) references to the data of Fig.~1.}
\begin{tabular}{@{}l c c c l  } 
\hline
Source &Other name &$z$ &Class. & Refs. for data \\
(1) & (2) &(3) & (4)  & (5) \\ 
\hline
0202$+$149 &4C 15.05  &0.833  &HPQ   &B85, BM94, C97, IT90, K81, NED, St92, St96, vM95\\
0208$-$512 &PKS       &1.003  &HPQ   &Be93, C97, IT88, NED, Sta96, To96     \\
0219$+$428 &3C 66A    &0.444  &LBL   &C97, Di96, G86, NED, Pi93, Si91, T95, WW90    \\
0234$-$285 &CTD 20    &1.213  &HPQ   &B95, BM91, E94, K81, NED, P82, St93, T92, vM95     \\
0235$+$164 &AO        &0.940  &LBL   &BM94, E94, G94, H93, K81, Ma96, NED, Pi93, S94, Si91, To96, WW90\\
0420$-$014 &PKS       &0.915  &HPQ   &BM94, Ch89, Co95, E94, IN88, K81, Li94, NED, Ra97, R95, S94, Si91, \\
           &          &       &      &Sm88, To96, WW90\\
0440$-$003 &NRAO 190  &0.844  &HPQ   &Bo90, K81, NED, St88, T95, To93, To96, W87, WP85    \\
0446$+$112 &PKS       &1.207  &NP    &NED, St88, T95, To96, W92   \\
0454$-$463 &PKS       &0.858  &LPQ   &B94, Fr83, IT90, K81, NED, To96, vM95, W91, WP85     \\
0521$-$365 &PKS       &0.055  &HPQ   &C97, IN88, NED, Pi93, Pi94, Pi96, T95, To96     \\
0528$+$134 &OG 147    &2.07   &LPQ   &B85, BM94, Co77, Co96, E94, IT90, Mc95, Mu96, NED, Po96, R82, R97, \\
           &          &       &      &Sa97, V92, WP85  \\
0537$-$441 &PKS       &0.896  &LBL   &Be92, C97, IN88, K81, L96, Ma85, NED, Pi93, Sa94, WP85, WW90  \\
0716$+$714 &S5        &$>$0.3 &LBL   &BM94, C94, E94, G97, IN88, K81, L95, NED, St92, W96  \\
0735$+$178 &PKS     &$>$0.424 &LBL   &C97, E94, G94, IN88, NED, N96, Pi93, Si91, To96, WW90 \\
0804$+$499 &0J 508    &1.433  &HPQ   &BM94, C97, NED, St94, vM95    \\
0805$-$077 &PKS       &1.837  &NP    &K81, NED, St93, T95  \\
0827$+$243 &OJ 248    &2.05   &LPQ   &B95, BM94, NED, Ra97, To96, V97, vM95    \\
0836$+$710 &4C 71.07  &2.172  &LPQ   &BM94, C97, E92, E94, K81, NED, Ra97, T93, W90, W92, WP85    \\
0917$+$449 &S4        &2.18   &LPQ   &BM94, C97, E94, NED, St93, T95  \\
0954$+$556 &4C55.17   &0.901  &HPQ   &BM94, C97, G94, K81, NED, Sr96   \\
0954$+$658 &S4        &0.368  &LBL   &C97, G94, IN88, K81, L86, Mu95, NED, St88   \\
1101$+$384 &Mkn 421   &0.031  &HBL   &Ma95, Ma96b     \\
1127$-$145 &PKS       &1.187  &LPQ   &A85, B94, Bo90, E94, IT90, K81, NED, Sr96, To96    \\
1156$+$295 &4C 29.45  &0.729  &HPQ   &BM94, E94, G94, IN88, Le85, Li94, LT96, NED, Pi93, Ra97, S94, Sa94, \\
           &          &       &      &Si91, To96, V97, vM95, We95    \\
1219$+$285 &ON 231    &0.102  &LBL   &BM91, C97, E92, E94, IN88, Lo90, NED, Pi93, Si91, To97, vM95, WW90     \\
1222$+$216 &4C 21.35  &0.435  &LPQ   &B95, NED, Sr96, To96    \\
1226$+$023 &3C 273    &0.158  &LPQ   &Al85, C83, G94, IN88, K81, L83, Mc95, NED, R86, Sh94, T90     \\
1229$-$021 &PKS       &1.045  &LPQ   &K81, NED, P88, Ra97, Sr96, St88, St94, To96, Wi94    \\
1253$-$055 &3C 279    &0.538  &HPQ   &Ma94     \\
1313$-$333 &PKS       &1.210  &NP    &N96, NED, St88, St92, St94, To96    \\
1406$-$076 &PKS       &1.494  &LPQ   &NED, St88, To93, To96, W95     \\
1424$-$418 &PKS       &1.522  &HPQ   &G81, IN88, NED, To96, T96 \\
1510$-$089 &PKS       &0.361  &HPQ   &C97, E94, G81, G94, IN88, K81, L86, LT97, NED, Pi93, Ra97, Sa94, Si91, \\
           &          &       &      &Sm88, Sr96, St93, To96, V97, WP85\\
\hline
\end{tabular}
\medskip 
\end{table*}

\setcounter{table}{0}
\begin{table*}
\caption[h]{{\it continue}}
\begin{tabular}{@{}l c c c l  } \hline
Source &Other name &$z$ &Class. & Refs. for data \\
(1) & (2) &(3) & (4)  & (5) \\ \hline
1604$+$159 &4C15.54   &0.357  &LBL   &GAM95, IN88, Le85, NED, Sr96   \\
1606$+$106 &4C 10.45  &1.227  &LPQ   &B95, Bi87, BM94, E94, IT90, K81, NED, To96, vM95    \\
1611$+$343 &DA 406    &1.404  &LPQ   &C97, E94, G94, K81, NED, Ra97, vM95     \\
1622$-$253 &PKS       &0.786  &LPQ   &N96, NED, St94    \\
1622$-$297 &PKS       &0.815  &LPQ   &K81, M97, NED, St93     \\
1633$+$382 &4C 38.41  &1.814  &LPQ   &BM91, BM94, Bo90, C97, IN88, K81, M93, Ra97, V92, V97, WP85  \\
1652$+$398 &Mkn 501   &0.055  &HBL   &B97, BM91, C97, G94, IN88, L91, Pi93, Q96, Sa94, St88, W92, WW90  \\
1730$-$130 &NRAO 530  &0.902  &NP    &B94, BM91, E94, N96, NED, St88, St93, To96, W88     \\
1739$+$522 &4C 51.37  &1.375  &HPQ   &B95, BM91, K81, NED, St88, St93, St94, V92, vM95    \\ 
1741$-$038 &OT-68     &1.054  &HPQ   &B94, E94, G94, K81, NED, St88, St92, St93, St94, To96, vM95   \\
1933$-$400 &PKS       &0.966  &NP    &B94, Di96, K81, NED, St94, To96   \\
2032$+$107 &PKS       &0.601  &LBL   &Di96, GAM95, NED, WW90  \\
2052$-$474 &PKS       &0.071  &LPQ   &B94, IT90, K81, NED, To96, vM95     \\
2155$-$304 &PKS       &0.117  &HBL   &U97, V96    \\
2200$+$420 &BL Lac    &0.069  &LBL   &BM94, Ca97, E94, IN88, NED, P96, Pi93, S94, St93, To96, U96    \\
2230$+$114 &CTA 102   &1.037  &HPQ   &B94, BM94, E94, F94, IN88, K81, Le85, Mc95, N93, Ne96, Ra97, St93, \\
           &          &       &      &T89, To96, W92, Wi94    \\
2251$+$158 &3C 454.3  &0.859  &HPQ   &Be92, BM91, BM94, C97, E94, G94, Ha93, IN88, K81, Le85, Mc95, \\
           &          &       &      &Ne96, NED, Pi93, Ra97, S94, Sm88 \\
2344$+$514 &1ES       &0.044  &HBL   &Cat97, NED, P96   \\
\hline
\end{tabular}
\medskip 
\end{table*}

\setcounter{table}{0}
\begin{table*}
\caption[h]{{\it continue}}
\begin{tabular}{@{}l l l l }
A85:& Adam (1985)  & 
Al85:& Aller et al. (1985)\\
B85:&  Bregman et al. (1985)& 
B94:& Brinkmann, Siebert \& Boller (1994)\\
B95:& Brinkmann et al. (1995)& 
B97:& Breslin et al. (1997)\\
Be92:& Bersanelli et al. (1992)& 
Be93:& Bertsch et al. (1993)\\
Bi87:& Biermann et al. (1987)&
BM91:& Bloom \& Marscher (1991)\\
BM94:& Bloom et al. (1994);&
Bo90:& Bozyan, Hemenway \& Argue (1990);\\
C83:& Clegg et al. (1983)&
C94:& Cappi et al. (1994)\\
C97:&  Comastri et al. (1997)&
Ca97:& Catanese et al. (1997a)\\
Cat97:& Catanese et al. (1997b)&
Ch89:& Chini, Biermann \& Gemund (1989)\\
Co77:& Condon, Hicks \& Jauncey (1977) &
Co95:& Condon, Anderson \& Broderick (1995)\\
Co96:& Collmar (1996)&
Di96:& Dingus et al. (1996)\\
E92:& Elvis et al. (1992)&
E94:&  Edelson (1994)\\
F94:& Falomo, Scarpa \& Bersanelli (1994)&
Fr83:& Friecke, Kollatschny \& Witzel  (1983)\\
GAM95:& Giommi, Ansari \& Micol (1995):&
G81:& Glass  (1981)\\
G86:&  Ghisellini et al. (1986)&
G94:&  Gear et al. (1994)\\
G97:&  Ghisellini et al. (1997)&
H93:&  Hunter et al. (1993)\\
Ha93:& Hartmann et al. (1993)&
IN88:& Impey \& Neugebauer (1988)\\
IT88:& Impey \& Tapia (1988)&
IT90:& Impey \& Tapia (1990)\\
K81:&  K\"uhr et al. (1981)&
L83:& Landau et al. (1983)\\
L86:& Landau et al. (1986)&
L91:& Lawrence et al. (1991)\\
L95:& Lin et al. (1995)&
L96:& Lin et al. (1996)\\
Le85:& Ledden \& O'Dell (1985)&
Li94:& Litchfield, Robson \& Stevens (1994)\\
Lo90:&  Lorenzetti et al. (1990)&
LT97:& Lawson \& Turner (1997)\\
M93:&  Mattox et al. (1993)&
M97:& Mattox et al. (1997)\\
Ma85:&  Maraschi et al. (1985)&
Ma94:& Maraschi et al. (1994)\\
Ma95:& Macomb et al. (1995)&
Ma96:& Madejski et al. (1996)\\
Ma96b:& Macomb et al. (1996)&
Mc95:& McNaron-Brown et al. (1995)\\
Mu95:& Mukherjee et al. (1995)&
Mu96:& Mukherjee et al. (1996)\\
N93:& Nolan et al. (1993)&
N96:& Nolan et al. (1996)\\
Ne96:& Netzer et al. (1996)&
P82:&  Perley (1982)\\
P88:& Pica et al. (1988)&
P96:& Perlman et al. (1996)\\
Pi93:& Pian et al. (1993)&
Pi94:& Pian et al. (1994)\\
Pi96:& Pian et al. (1996)&
Po96:& Pohl et al. (1996)\\
Q96:&  Quinn et al. (1996)&
R82:&  Rieke, Lebofsky \& Wi\'sniewski (1982)\\
R86:&  Robson  et al. (1986)&
R95:&  Radecke et al. (1995)\\
R97:&  Reuter et al. (1997)&
Ra97:& Raiteri et al. (1997)\\
S94:&  Stevens et al. (1994)&
Sa94:& Sambruna et al. (1994)\\
Sa97:& Sambruna et al. (1997)&
Sh94:& Schonfelder (1994)\\
Si91:& Sitko \& Sitko (1991)&
Sm88:& Smith et al. (1988)\\
Sr96:& Sreekumar et al. (1996)& 
St88:& Steppe et al. (1988)\\
St92:& Steppe et al. (1992)&
St93:& Steppe et al. (1993)\\
St94:& Stickel, Meisenheimer \& Kuhr (1994)&
St96:& Stickel et al. (1996)\\
Sta96:& Stacy et al.  (1996)&
T89:&  Terasranta et al. (1989)\\
T90:&  Turner et al. (1990)&
T92:&  Terasranta et al. (1992)\\
T93:&  Thompson et al. (1993)&
T95:&  Thompson et al. (1995)\\
T96:&  Thompson et al. (1996) &
To93:& Tornikoski et al. (1993)\\
To96:& Tornikoski et al. (1996)&
To97:& Tosti et al. (1997)\\
U96:&  Urry et al. (1996)&
U97:&  Urry et al. (1997)\\
V92:&  Valtaoja, Lahteenmaki \& Terasranta (1992)&
V96:&  Vestrand et al. (1996)\\
V97:&  Villata  et al. (1997)&
vM95:& von Montigny et al. (1995)\\
W87:&  Worral et al. (1987)&
W88:&  Webb et al. (1988)\\
W90:&  Wagner et al. (1990)&
W91:&  Wright et al. (1991) \\
W92:&  Wiren et al. (1992)&
W95:&  Wagner et al. (1995)\\
W96:&  Wagner et al. (1996)&
We95:& Webb et al. (1995)\\
Wi94:& Wilkes et al. (1994)&
WP85:& Wall \& Peacock (1985)\\
WW90:& Worral \& Wilkes (1990)\\
\end{tabular}
\end{table*}

\begin{table*}
\caption{The input parameters for the EC and SSC models are reported
in the first and second line for each source,
respectively.  (1) Source name; (2) region size in units of $10^{15}$
cm; (3), (4) compactnesses in injected particles and external
radiation field; (5) maximum energy of the injected particles; (6)
energy of the peak of the stationary electron distribution; (7)
spectral index of the injected particles; (8) magnetic field intensity
(in Gauss); (9) relativistic Doppler factor. The SSC model for
1253--055 requires monoenergetic injection.}
\begin{tabular}{@{}lcccccccc}
\hline Source      &$R/10^{15}$  &$\ell_{\rm inj}$ &$\ell_{\rm ext}$
&$\gamma_{\rm max}$ 
&$\gamma_{\rm peak}$  &$s$   &$B$   &$\delta$\\
(1) & (2) &(3) & (4) & (5)& (6) & (7) &(8) & (9) \\ 
\hline

0202$+$149  &30 &0.05   &2     &3.0e3  &2.5e2  &3.0 &3.040 &14     \\
            &80 &0.01   &--    &4.0e4  &1.0e4  &3.3 &0.026 &18    \\
0208$-$512  &70 &0.02   &0.06  &7.0e3  &1.0e3  &3.8 &0.563 &23     \\
            &50 &0.08   &--    &3.0e4  &1.5e4  &2.0 &0.106 &16     \\
0219$+$428  &20 &0.03   &3e-3  &1.0e5  &8.0e3  &2.4 &2.040 &13.5  \\
            &40 &0.01   &--    &2.0e5  &1.0e4  &2.4 &0.590 &15     \\
0234$-$285  &40 &0.06   &1     &1.0e4  &2.0e2  &2.9 &4.080 &16     \\
            &50 &0.08   &--    &6.0e4  &1.0e4  &3.0 &0.067 &13     \\
0235$+$164  &50 &0.05   &0.03  &8.0e4  &3.0e3  &3.0 &0.912 &20     \\
            &60 &0.02   &--    &3.0e5  &1.0e4  &2.9 &0.215 &21   \\
0420$-$014  &50 &0.03   &0.04  &1.0e4  &2.0e3  &3.5 &0.745 &16     \\
            &80 &0.01   &--    &8.0e4  &1.5e4  &3.0 &0.034 &20     \\
0440$-$003  &50 &0.025  &0.025 &2.0e4  &2.0e3  &3.0 &0.680 &17     \\
            &70 &0.01   &--    &8.0e4  &2.0e4  &3.0 &0.106 &20     \\
0446$+$112  &50 &0.08   &1     &4.0e3  &1.5e2  &2.0 &0.810  &20     \\
            &90 &0.08   &--    &1.0e5  &1.0e4  &3.5 &0.011 &20     \\
0454$-$463  &30 &0.03   &0.08  &8.0e3  &6.0e2  &2.1 &1.050 &16     \\
            &80 &7e-3   &--    &8.0e4  &3.0e4  &2.5 &0.018 &20     \\
0521$-$365  &50 &1.00   &0.10  &5.0e4  &2.0e3  &2.5 &3.141 &1.4    \\
            &50 &0.60   &--    &7.0e4  &3.5e3  &2.7 &2.580 &1.6    \\
0528$+$134  &65 &0.90   &7     &6.0e3  &3.0e2  &2.6 &6.198 &15     \\
            &60 &0.60   &--    &8.0e4  &3.0e3  &3.0 &0.215 &19     \\
0537$-$441  &70 &0.04   &1e-3  &7.0e4  &4.0e3  &2.2 &0.325 &15     \\
            &50 &0.05   &--    &1.0e5  &4.0e3  &2.5 &0.430 &15     \\
0716$+$714  &30 &0.02   &7e-3  &3.0e4  &2.0e3  &2.6 &1.813 &11.5   \\
            &50 &3e-3   &--    &5.0e4  &4.0e3  &2.7 &0.460 &15     \\
0735$+$178  &40 &0.01   &4e-3  &3.0e4  &3.0e3  &3.0 &0.833 &14   \\
            &50 &3e-3   &--    &4.0e4  &5.0e3  &3.3 &0.408 &17     \\
0804$+$499  &50 &0.10   &0.70  &8.0e3  &4.0e2  &3.1 &3.333 &15     \\
            &70 &0.06   &--    &8.0e4  &1.0e4  &3.4 &0.073 &17     \\
0805$-$077  &40 &0.20   &2     &6.0e3  &4.0e2  &2.5 &2.483 &17     \\
            &90 &0.09   &--    &1.0e5  &9.0e3  &3.7 &0.037 &20     \\
0827$+$234  &50 &0.20   &0.90  &8.0e3  &3.0e2  &2.3 &4.711 &16     \\
            &70 &0.20   &--    &8.0e4  &5.0e3  &2.7 &0.563 &16     \\
0836$+$710  &50 &0.70   &6.00  &7.0e3  &1.8e2  &3.1 &8.814 &18     \\
            &70 &0.30   &--    &5.0e4  &4.0e3  &3.0 &0.282 &17     \\
0917$+$449  &30 &0.50   &8     &6.0e3  &2.0e2  &2.1 &2.267 &13     \\
            &90 &0.03   &--    &1.0e5  &4.0e3  &2.3 &0.025 &23     \\
0954$+$556  &50 &0.01   &0.02  &1.5e4  &5.0e3  &1.7 &0.527 &15     \\
            &70 &4e-3   &--    &7.0e4  &1.0e4  &0.0  &0.056 &17     \\
0954$+$658  &30 &4.5e-3 &0.025 &6.0e3  &9.0e2  &3.6 &2.040 &13     \\
            &60 &7e-4   &--    &7.0e4  &2.0e4  &2.0  &0.025 &18     \\
1101$+$384  &5  &6e-3   &1e-3  &8.0e5  &6.0e4  &2.0 &0.222 &11     \\ 
            &10 &2e-3   &--    &8.0e5  &2.0e5  &1.2 &0.093 &12     \\
1127$-$145  &60 &0.15   &0.80  &8.0e3  &3.0e2  &2.3 &1.862 &15.5   \\
            &70 &0.10   &--    &8.0e4  &1.5e4  &2.3 &0.048 &17     \\
1156$+$295  &20 &0.15   &0.08  &1.0e4  &3.0e3  &2.4 &1.360 &15     \\
            &70 &0.02   &--    &8.0e4  &9.0e3  &2.0  &0.252 &18     \\
1219$+$285  &20 &1.e-3 &4.e-3  &1.0e5  &5.0e3  &4.2 &1.178 &11     \\
            &70 &5.e-5  &--    &7.0e4  &4.0e4  &0.0  &0.036 &20     \\
1222$+$216  &10 &0.10   &0.25  &8.0e3  &3.5e2  &2.9 &3.333 &11     \\
            &40 &0.02   &--    &4.0e4  &9.0e3  &3.2 &0.068 &11     \\
1226$+$023  &10 &1.00   &1.50  &1.0e4  &8.0e1  &3.2 &8.900 &6.5    \\
            &40 &0.06   &--    &3.0e4  &3.5e3  &3.2 &0.456 &7.0     \\
\end{tabular}
\end{table*}

\setcounter{table}{1}
\begin{table*}
\caption{{\it continue}}
\begin{tabular}{@{}lcccccccc} \hline
Source      &$R/10^{15}$  &$\ell_{\rm inj}$ &$\ell_{\rm ext}$
&$\gamma_{\rm max}$ 
&$\gamma_{\rm peak}$  &$s$   &$B$   &$\delta$ \\
(1) & (2) &(3) & (4) & (5)& (6) & (7) &(8) & (9) 
\\ \hline
1229$-$021  &40 &0.08   &1.00  &8.0e3  &3.0e2  &3.8 &2.356 &12     \\
            &50 &0.03   &--    &2.0e4  &1.0e4  &4.0 &0.105 &16     \\
1253$-$055  &30 &0.04   &0.07  &7.0e3  &8.0e2  &3.2 &1.360 &18     \\
            &80 &0.04   &--    &2.0e4  &2.0e4  &--  &0.059 &15     \\
1313$-$333  &40 &0.03   &0.70  &9.0e3  &1.0e3  &1.9 &1.180 &17     \\
            &70 &0.01   &--    &1.0e5  &3.0e4  &1.0 &0.009 &20     \\
1406$-$076  &60 &0.05   &0.10  &1.0e4  &3.0e3  &1.5 &1.080 &21     \\
            &70 &0.08   &--    &6.0e4  &1.5e4  &2.3 &0.206 &18     \\
1424$-$418  &30 &0.50   &1.00  &6.0e3  &2.0e2  &2.7 &4.060 &15     \\
            &90 &0.07   &--    &1.0e5  &5.0e3  &3.0 &0.093 &20     \\
1510$-$089  &20 &0.05   &0.80  &1.0e4  &1.2e2  &3.3 &5.890 &13     \\
            &30 &4e-3   &--    &3.0e4  &4.0e3  &2.2 &0.061 &18   \\
1604$+$159  &20 &0.01   &0.10  &1.0e4  &9.0e2  &2.1 &0.960 &15     \\
            &50 &2e-3   &--    &7.0e4  &3.0e4  &2.1 &0.011 &18     \\
1606$+$106  &30 &0.30   &3.00  &5.0e3  &1.5e2  &3.0 &3.330 &15     \\
            &70 &0.06   &--    &3.0e4  &1.0e4  &3.0 &0.028 &18     \\
1611$+$343  &30 &0.20   &1.00  &5.0e3  &2.0e2  &2.6 &3.041 &16     \\
            &50 &0.20   &--    &6.0e4  &1.3e4  &3.3 &0.105 &13.5   \\
1622$-$253  &30 &0.03   &0.30  &6.0e3  &3.0e2  &2.5 &1.670 &15     \\
            &70 &0.01   &--    &8.0e4  &2.0e4  &2.3 &0.013 &16     \\
1622$-$297  &20 &0.10   &1.00  &2.5e3  &6.0e2  &1.5 &2.150 &23     \\
            &80 &0.03   &--    &5.0e4  &3.0e4  &2.1 &0.007 &21     \\
1633$+$382  &60 &0.18   &0.60  &6.0e3  &1.5e3  &1.5 &1.667 &21     \\
            &80 &0.20   &--    &7.0e4  &2.0e4  &2.0 &0.052 &19     \\
1652$+$398  &5  &2e-3   &5e-4  &8.0e5  &1.0e4  &2.8 &1.110 &10     \\
            &10 &1e-3   &--    &8.0e5  &2.0e4  &3.0 &0.497 &10     \\
1730$-$130  &30 &0.04   &0.40  &3.0e4  &4.0e2  &2.8 &5.440 &17     \\
            &60 &0.02   &--    &6.0e4  &6.0e3  &2.4 &0.192 &16     \\
1739$+$522  &60 &0.03   &0.08  &1.0e4  &4.0e2  &2.2 &0.450 &20     \\
            &70 &0.04   &--    &1.0e5  &2.0e4  &2.1 &0.019 &19     \\
1741$-$038  &60 &0.05   &0.90  &8.0e3  &2.5e2  &3.8 &2.150 &17     \\
            &60 &0.05   &--    &2.0e4  &1.0e4  &3.0 &0.048 &17.5   \\
1933$-$400  &20 &0.07   &1.00  &6.0e3  &3.0e2  &2.6 &4.410 &14     \\
            &50 &0.03   &--    &5.0e4  &8.0e3  &2.9 &0.060 &14     \\
2032$+$107  &20 &0.06   &0.20  &6.0e3  &3.0e2  &3.0 &3.850 &12     \\
            &50 &5e-3   &--    &5.0e4  &4.0e3  &3.3 &0.118 &20     \\
2052$-$474  &50 &0.20   &2.00  &7.0e3  &2.0e2  &2.8 &1.920 &15     \\
            &70 &0.10   &--    &5.0e4  &8.0e3  &2.9 &0.073 &16     \\
2155$-$304  &20 &2e-3   &3e-4  &4.0e5  &8.0e3  &2.4 &1.050 &17     \\
            &20 &2e-3   &--    &1.0e6  &7.0e3  &2.6 &1.216 &18     \\
2200$+$420  &8  &8e-3   &3e-4  &3.0e5  &2.7e3  &2.8 &1.670 &10     \\   
            &20 &2e-3   &--    &3.0e5  &5.0e3  &2.8 &0.430 &11     \\
2230$+$114  &40 &0.80   &5.00  &1.0e4  &1.0e2  &3.1 &8.600 &10     \\
            &70 &0.03   &--    &3.0e4  &6.0e3  &2.9 &0.077 &18     \\
2251$+$158  &40 &0.80   &5.00  &6.0e3  &1.0e2  &2.3 &7.450 &10     \\
            &70 &0.04   &--    &6.0e4  &4.0e3  &2.2 &0.073 &18     \\
2344$+$512  &8  &1.4e-4 &1e-4  &7.0e5  &4.0e4  &3.7 &0.470 &14     \\   
            &10 &2e-4   &--    &8.0e5  &4.5e4  &3.5 &0.220 &13     \\
\hline
\end{tabular}
\end{table*}

\begin{table*}
\caption{Linear correlations for the EC model. (1), (2), (4), (5)
parameters of the correlation of the form $y= m x + q$; (3) number of
objects; (6) correlation coefficient; (7) probability of a random
distribution; (8) sources considered.}
\begin{tabular}{@{}llcrrccl} \hline
$y$ &$x$ &$N$  &$\qquad m$ &$\qquad q$ &$\, \, r$ &$\quad P$ &Objects\\
(1) & (2) &(3) & (4) & (5)& (6) & (7) &(8) \\ 
\hline
$\log \gamma_{\rm peak}$   &$\log (U_{\rm r}+U_{\rm B})$
  &51 &$ -0.63\pm 0.04$ & $2.97\pm 0.04$ &0.902 &$7.2\times 10^{-10}$ &All  \\
& &37 &$ -0.50\pm 0.06$ & $2.88\pm 0.05$ &0.812 &$2.5\times 10^{-9}$  &Only FSRQ \\
& &14 &$ -0.80\pm 0.12$ & $2.88\pm 0.14$ &0.886 &$2.4\times 10^{-5}$  &Only BL Lacs \\
$\log \gamma_{\rm peak}$  &$\log U_{\rm B}$      
  &51 &$ -0.64\pm 0.09$ & $2.33\pm 0.10$ &0.735 &$2.3\times 10^{-9}$ &All      \\
& &37 &$ -0.43\pm 0.08$ & $2.33\pm 0.08$ &0.676 &$4.5\times 10^{-6}$ &Only FSRQ \\
& &14 &$ -0.67\pm 0.19$ & $2.70\pm 0.28$ &0.719 &$3.7\times 10^{-3}$ &Only BL Lacs \\
$\log \gamma_{\rm peak}$  &$\log U_{\rm r}$    
  &51 &$ -0.56\pm 0.04$ & $2.89\pm 0.04$ &0.911 &$1.2\times 10^{-10}$ &All      \\
& &37 &$ -0.50\pm 0.06$ & $2.85\pm 0.05$ &0.818 &$7.9\times 10^{-11}$ &Only FSRQ \\
& &14 &$ -0.64\pm 0.10$ & $2.82\pm 0.15$ &0.875 &$4.0\times 10^{-5}$  &Only BL Lacs \\
$\log \gamma_{\rm peak}$  &$\log \ell_{\rm ext}$ 
  &51 &$ -0.48\pm 0.04$ & $2.47\pm 0.05$ &0.890 &$7.1\times 10^{-10}$ &All      \\
& &37 &$ -0.54\pm 0.07$ & $2.47\pm 0.05$ &0.799 &$1.6\times 10^{-9}$  &Only FSRQ \\
& &14 &$ -0.49\pm 0.11$ & $2.39\pm 0.29$ &0.794 &$7.0\times 10^{-4}$  &Only BL Lacs \\
$\log \gamma_{\rm peak}$  &$\log \ell_{\rm inj}$ 
  &51 &$ -0.61\pm 0.08$ & $2.11\pm 0.12$ &0.735 &$3.4\times 10^{-9}$  &All      \\
& &37 &$ -0.41\pm 0.12$ & $2.23\pm 0.14$ &0.491 &$2.0\times 10^{-3}$  &Only FSRQ \\
& &14 &$ -0.48\pm 0.21$ & $2.59\pm 0.48$ &0.566 &$3.5\times 10^{-2}$  &Only BL Lacs \\
$\log \gamma_{\rm peak}$  &$\log (\ell_{\rm inj}+\ell_{\rm ext})$ 
  &51 &$ -0.58\pm 0.05$ & $2.54\pm 0.06$ &0.871 &$1.7\times 10^{-10}$ &All      \\
& &37 &$ -0.55\pm 0.09$ & $2.54\pm 0.05$ &0.740 &$1.7\times 10^{-7}$  &Only FSRQ \\
& &14 &$ -0.61\pm 0.14$ & $2.49\pm 0.29$ &0.776 &$1.1\times 10^{-3}$  &Only BL Lacs \\
$\log \gamma_{\rm peak}$  &$\log L_{\rm C}/L_{\rm syn }$
  &51 &$ -0.70\pm 0.11$ & $3.53\pm 0.12$ &0.690 &$2.2\times 10^{-8}$  &All      \\
& &37 &$ -0.44\pm 0.18$ & $3.15\pm 0.23$ &0.381 &$2.0\times 10^{-2}$  &Only FSRQ\\
& &14 &$ -0.46\pm 0.28$ & $3.67\pm 0.16$ &0.427 &$1.3\times 10^{-1}$  &Only BL Lacs \\
 & & & & & & & \\
$\log L_{\rm C}/L_{\rm syn }$ &$\log \ell_{\rm inj}$ 
  &51 & $0.49\pm 0.09$ & $1.54\pm 0.14$ &0.601 &$2.3\times 10^{-6}$  &All      \\
& &37 & $0.02\pm 0.12$ & $1.18\pm 0.13$ &0.027 &$8.7\times 10^{-1}$  &Only FSRQ\\
& &14 & $0.47\pm 0.18$ & $1.15\pm 0.41$ &0.599 &$2.4\times 10^{-2}$  &Only BL Lacs \\
$\log L_{\rm C}/L_{\rm syn }$ &$\log \ell_{\rm ext}$ 
  &51 & $0.42\pm 0.05$ & $1.28\pm 0.07$ &0.797 &$2.2\times 10^{-9}$ &All      \\
& &37 & $0.25\pm 0.09$ & $1.27\pm 0.07$ &0.426 &$8.6\times 10^{-3}$  &Only FSRQ \\
& &14 & $0.38\pm 0.13$ & $1.07\pm 0.34$ &0.649 &$1.2\times 10^{-2}$  &Only BL Lacs \\
$\log L_{\rm C}/L_{\rm syn }$  &$\log (\ell_{\rm inj}+\ell_{\rm ext})$ 
  &51 & $0.50\pm 0.06$ & $1.22\pm 0.07$ &0.771 &$6.7\times 10^{-10}$  &All      \\
& &37 & $0.20\pm 0.10$ & $1.23\pm 0.07$ &0.311 &$6.1\times 10^{-2}$   &Only FSRQ \\
& &14 & $0.55\pm 0.14$ & $1.15\pm 0.28$ &0.749 &$2.0\times 10^{-3}$   &Only BL Lacs \\
$\log L_{\rm C}/L_{\rm syn }$  &$\log L_{\rm inj}^{\rm obs}$ 
  &51 & $0.40\pm 0.06$ & $-18.44\pm 2.81$ &0.702 &$8.7\times 10^{-9}$ &All      \\
& &37 & $0.17\pm 0.08$ & $-7.28\pm 4.03$  &0.335 &$4.2\times 10^{-2}$ &Only FSRQ \\
& &14 & $0.29\pm 0.12$ & $-13.44\pm 5.61$ &0.573 &$3.2\times 10^{-2}$ &Only BL Lacs \\
$\log L_{\rm C}/L_{\rm syn }$  &$\log \nu_{\rm peak}^{\rm obs}$ 
  &51 & $-0.45\pm 0.06$ & $11.57\pm 1.40$ &0.736 &$1.4\times 10^{-9}$  &All      \\
& &37 & $-0.25\pm 0.08$ & $6.92\pm 1.93$  &0.449 &$5.3\times 10^{-3}$  &Only FSRQ \\
& &14 & $-0.31\pm 0.14$ & $7.88\pm 3.45$  &0.544 &$4.4\times 10^{-2}$  &Only BL Lacs \\
$\log L_{\rm C}/L_{\rm syn }$  &$\log B$
  &51 & $ 0.37\pm 0.24$ & $0.81\pm 0.11$ &0.217 &$1.3\times 10^{-1}$ &All      \\
& &37 & $-0.15\pm 0.19$ & $1.26\pm 0.09$ &0.132 &$4.3\times 10^{-1}$ &Only FSRQ \\
& &14 & $-0.25\pm 0.50$ & $0.14\pm 0.16$ &0.145 &$6.2\times 10^{-1}$ &Only BL Lacs \\
 & & & & & & & \\
$\log \ell_{\rm ext}$ &$\log \ell_{\rm inj}$ 
  &51 & $1.28\pm 0.13$ & $0.76\pm 0.19$ &0.822 &$6.3\times 10^{-10}$  &All      \\
& &37 & $0.83\pm 0.16$ & $0.52\pm 0.17$ &0.671 &$5.5\times 10^{-6}$   &Only FSRQ\\
& &14 & $0.82\pm 0.31$ &$-0.71\pm 0.71$ &0.605 &$2.2\times 10^{-2}$  &Only BL Lacs \\
 & & & & & & & \\
$\log U_B$  &$\log U_r$  
  &51 & $0.57\pm 0.06$ & $-0.86\pm 0.06$ &0.812 &$1.2\times 10^{-10}$  &All      \\
& &37 & $0.82\pm 0.08$ & $-1.05\pm 0.07$ &0.857 &$5.7\times 10^{-10}$ &Only FSRQ \\
& &14 & $0.54\pm 0.16$ & $-0.69\pm 0.24$ &0.693 &$6.0\times 10^{-3}$  &Only BL Lacs \\
\hline
\end{tabular}
\end{table*}

\begin{table*}
\caption{Linear correlations for the SSC model. (1), (2), (4), (5)
parameters of the correlation of the form $y= m x + q$; (3) number of
objects; (6) correlation coefficient; (7) probability of a random
distribution; (8) sources considered.}
\begin{tabular}{@{}llcrrccl} \hline
$y$ &$x$ &$N$  &$\qquad m$ &$\qquad q$ &$\, \, r$ &$\quad P$ &Objects\\
(1) & (2) &(3) & (4) & (5)& (6) & (7) &(8) \\ 
\hline
$\log \gamma_{\rm peak}$  &$\log (U_{\rm r}+U_{\rm B})$    
  &51  &$ -0.29\pm 0.05$ & $3.37\pm 0.13$ &0.616 &$1.5\times 10^{-6}$ &All      \\
& &37  &$ -0.24\pm 0.05$ & $3.43\pm 0.12$ &0.627 &$3.3\times 10^{-5}$ &Only FSRQ \\
& &14  &$ -0.36\pm 0.13$ & $3.29\pm 0.31$ &0.638 &$1.4\times 10^{-2}$ &Only BL Lacs \\
$\log \gamma_{\rm peak}$  &$\log U_{\rm B}$        
  &51  &$ -0.15\pm 0.04$ & $3.51\pm 0.15$ &0.466 &$5.7\times 10^{-4}$  &All      \\
& &37  &$ -0.17\pm 0.04$ & $3.35\pm 0.14$ &0.627 &$3.2\times 10^{-5}$  &Only FSRQ \\
& &14  &$ -0.23\pm 0.11$ & $3.48\pm 0.32$ &0.524 &$5.4\times 10^{-2}$  &Only BL Lacs \\
$\log \gamma_{\rm peak}$  &$\log U_{\rm r}$    
  &51  &$ -0.31\pm 0.05$ & $3.29\pm 0.13$ &0.642 &$3.7\times 10^{-7}$  &All      \\
& &37  &$ -0.25\pm 0.05$ & $3.41\pm 0.13$ &0.627 &$3.3\times 10^{-5}$  &Only FSRQ \\
& &14  &$ -0.41\pm 0.13$ & $3.07\pm 0.35$ &0.672 &$8.5\times 10^{-3}$  &Only BL Lacs \\
$\log \gamma_{\rm peak}$  &$\log \ell_{\rm inj}$ 
  &51  &$ -0.19\pm 0.06$ & $3.70\pm 0.11$ &0.434 &$1.5\times 10^{-3}$  &All      \\
& &37  &$ -0.19\pm 0.08$ & $3.72\pm 0.12$ &0.365 &$2.6\times 10^{-2}$  &Only FSRQ \\
& &14  &$ -0.36\pm 0.16$ & $3.16\pm 0.45$ &0.539 &$4.7\times 10^{-2}$  &Only BL Lacs \\
$\log \gamma_{\rm peak}$ &$\log L_{\rm C}/L_{\rm syn}$     
  &51  &$  0.05\pm 0.07$ & $3.96\pm 0.10$ &0.100 &$4.8\times 10^{-1}$  &All      \\
& &37  &$  0.32\pm 0.09$ & $3.52\pm 0.14$ &0.504 &$1.5\times 10^{-3}$  &Only FSRQ \\
& &14  &$  0.12\pm 0.23$ & $4.09\pm 0.16$ &0.143 &$6.2\times 10^{-1}$  &Only BL Lacs \\
 & & & & & & & \\
$\log L_{\rm C}/L_{\rm syn}$   &$\log \ell_{\rm inj}$
  &51  & $0.47\pm 0.10$  & $1.94\pm 0.19$ &0.555 &$2.4\times 10^{-5}$  &All      \\
& &37  &$-0.03\pm 0.14$  & $1.41\pm 0.20$ &0.031 &$8.6\times 10^{-1}$  &Only FSRQ \\
& &14  & $0.24\pm 0.23$  & $0.95\pm 0.64$ &0.292 &$3.1\times 10^{-1}$  &Only BL Lacs \\
$\log L_{\rm C}/L_{\rm syn }$  &$\log L_{\rm inj}^{\rm obs}$ 
  &51 & $0.45\pm 0.06$ & $-20.52\pm 2.94$ &0.725 &$5.9\times 10^{-9}$ &All      \\
& &37 & $0.27\pm 0.08$ & $-11.84\pm 3.90$ &0.499 &$1.7\times 10^{-3}$ &Only FSRQ \\
& &14 & $0.28\pm 0.15$ & $-13.00\pm 6.88$ &0.487 &$7.7\times 10^{-2}$ &Only BL Lacs \\
$\log L_{\rm C}/L_{\rm syn }$  &$\log \nu_{\rm peak}^{\rm obs}$ 
  &51 & $-0.54\pm 0.14$ & $14.57\pm 3.39$ &0.493 &$2.4\times 10^{-4}$  &All      \\
& &37 & $-0.15\pm 0.16$ & $5.19\pm 4.09$  &0.156 &$3.5\times 10^{-1}$  &Only FSRQ \\
& &14 & $-0.25\pm 0.21$ & $6.62\pm 5.35$  &0.324 &$2.6\times 10^{-1}$  &Only BL Lacs \\
$\log L_{\rm C}/L_{\rm syn }$  &$\log B$
  &51 & $-0.98\pm 0.11$ & $0.10\pm 0.13$  &0.783 &$1.2\times 10^{-9}$ &All      \\
& &37 & $-0.74\pm 0.07$ & $0.58\pm 0.09$  &0.864 &$1.5\times 10^{-11}$ &Only FSRQ \\
& &14 & $-0.82\pm 0.20$ & $-0.30\pm 0.19$ &0.761 &$1.6\times 10^{-3}$ &Only BL Lacs \\
 & & & & & & & \\
$\log U_B$  &$\log U_r$ 
  &51 & $1.21\pm 0.13$ & $-0.64\pm 0.33$ &0.791 &$2.3\times 10^{-9}$  &All      \\
& &37 & $1.39\pm 0.08$ & $-0.57\pm 0.19$ &0.945 &$6.3\times 10^{-11}$  &Only FSRQ \\
& &14 & $1.22\pm 0.21$ & $0.27\pm 0.57$  &0.857 &$9.0\times 10^{-5}$   &Only BL Lacs \\
\hline
\end{tabular}
\end{table*}

\begin{table*}
\caption{The input parameters for the EC model reproducing the average
SEDs determined by Fossati et al. (1998) by dividing into radio
luminosity (L$_{\rm R}$) bins BL Lacs and FSRQ belonging to complete
samples. (1) Radio luminosity bin; (2) region size in units of
$10^{15}$ cm; (3), (4) compactnesses in injected particles and
external radiation field; (5) maximum energy of the injected
particles; (6) energy of the peak of the stationary electron
distribution; (7) spectral index of the injected particles; (8)
magnetic field intensity (in Gauss). In the fits the relativistic
Doppler factor (column 9) has been fixed at the value $\delta$=15.}
\begin{tabular}{@{}lcccccccc}
\hline Log L$_{\rm R}$ &$R/10^{15}$  &$\ell_{\rm inj}$ &$\ell_{\rm ext}$
&$\gamma_{\rm max}$ 
&$\gamma_{\rm peak}$  &$s$   &$B$ & $\delta$  \\
(1) & (2) &(3) & (4) & (5)& (6) & (7) &(8)& (9) \\ 
\hline
41.5  & 10 & 7e-4 & 1e-5 &1.0e6 & 2.5e4 &3.0 &0.441 &15\\
42.5  & 10 & 1e-3 & 1e-3 &2.0e5 & 2.0e3 &2.8 &0.745 &15\\
43.5  & 30 & 5e-3 & 2e-2 &3.0e4 & 5.0e2 &3.0 &1.075 &15\\
44.5  & 40 & 5e-2 & 0.1  &1.5e4 & 3.5e2 &2.5 &1.178 &15\\
45.5  & 50 & 5e-1 & 5.0  &6.0e3 & 1.5e2 &2.8 &6.082 &15\\
\hline
\end{tabular}
\end{table*}

\begin{table*}
\caption{Results of the principal component analysis. EV stands
for eigenvector. The first row of the table lists the percentage of
the correlation accounted for by the different eigenvectors.}
\begin{tabular}{@{}lrrrrrr}
\hline
Variable            &EV$_1$    &EV$_2$     &EV$_3$    &EV$_4$    &EV$_5$    &EV$_6$  \\
\hline
\%                  &44.55     &27.08      &13.17     &9.0       &3.52      &2.66    \\
\hline
$\delta$            &$-$0.2408 &$-$0.5922  &$-$0.3525 & 0.5140   & 0.1872   &$-$0.4098  \\
$\ell_{\rm inj}$    &0.5608    & 0.0623    &$-$0.0525 &$-$0.3317 & 0.0915   &$-$0.7486   \\
$\ell_{\rm ext}$    &0.5242    &$-$0.1578  &$-$0.3110 & 0.2645   &$-$0.7031 &0.1982   \\
$\gamma_{\rm peak}$ &$-$0.2086 & 0.4119    &$-$0.8594 &$-$0.2063 & 0.0652   &0.0376  \\
$U_B$               &0.5510    & 0.0746    &$-$0.1082 & 0.3196   & 0.6654   &0.3663   \\
$R$                 &0.0748    &$-$0.6673  &$-$0.1611 &$-$0.6412 & 0.1234   &0.3111   \\
\hline
\end{tabular}
\end{table*}

\end{document}